\documentclass[]{spie}  
\usepackage[]{graphicx}

\usepackage{color}
\usepackage{bm}

\usepackage[english]{babel}
\usepackage{amsmath}
\usepackage{amssymb}
\usepackage{units}
\usepackage{upgreek}
\usepackage[colorlinks=true, allcolors=blue]{hyperref}
\usepackage{ctable}
\usepackage{footnote}
\usepackage{bbold}
\usepackage{textcomp}
\usepackage[absolute,overlay]{textpos}
\usepackage{helvet}

\title{Beam shaping for ultrafast materials processing} 

\author{Daniel Flamm,\supit{a} Daniel G\"{u}nther Gro{\ss}mann,\supit{b} Michael Jenne,\supit{a} Felix Zimmermann,\supit{a} Jonas Kleiner,\supit{a} Myriam Kaiser,\supit{a} Julian Hellstern,\supit{b} Christoph Tillkorn,\supit{b} and \mbox{Malte Kumkar\supit{a}}
\skiplinehalf
\supit{a}TRUMPF Laser- und Systemtechnik GmbH, Johann-Maus-Str.\,2, 71254 Ditzingen, Germany \\
\supit{b}TRUMPF Laser GmbH, Aichhalder Str.\,39, 78713 Schramberg, Germany 
}

\authorinfo{Further author information:\\ E-Mail: \href{mailto:daniel.flamm@trumpf.com}{daniel.flamm@trumpf.com}.}

\begin{document} 
  \maketitle 

\begin{abstract}
The remarkable temporal properties of ultra-short pulsed lasers in combination with novel beam shaping concepts enable the development of completely new material processing strategies. We demonstrate the benefit of employing focus distributions being tailored in all three spatial dimensions. As example advanced Bessel-like beam profiles, 3D-beam splitting concepts and flat-top focus distributions are used to achieve high-quality and efficient results for cutting, welding and drilling applications. Spatial and temporal in situ diagnostics is employed to analyze light-matter interaction and, in combination with flexible digital-holographic beam shaping techniques, to find the optimal beam shape for the respective laser application.
\end{abstract}



\section{INTRODUCTION}
\label{sec:intro}  

\begin{textblock*}{16cm}(2.67cm,1cm) 
   \centering
  \tiny \textsf{Daniel Flamm, Daniel Günther Grossmann, Michael Jenne, Felix
Zimmermann, Jonas Kleiner, Myriam Kaiser, Julian Hellstern, Christoph
Tillkorn, Malte Kumkar, "Beam shaping for\\ ultrafast materials processing,"
Proc. SPIE 10904, Laser Resonators, Microresonators, and Beam Control
XXI, 109041G (4 March 2019); \url{https://doi.org/10.1117/12.2511516}}
\end{textblock*}

\begin{textblock*}{5cm}(9.2cm,1.9cm) 
   \textsf{Invited Paper}
\end{textblock*}

\begin{textblock*}{17cm}(2.25cm,25.25cm) 
   \centering \small 
   \textsf{
   © 2019 Society of Photo‑Optical Instrumentation Engineers (SPIE). One print or electronic copy may be made for personal use only. Systematic reproduction and distribution, duplication of any material in this publication for a fee or for commercial purposes, and modification of the contents of the publication are prohibited. \\
   Laser Resonators, Microresonators, and Beam Control XXI, edited by Alexis V. Kudryashov, Alan H. Paxton,\\ Vladimir S. Ilchenko, Proc. of SPIE Vol. 10904, 109041G. © 2019 SPIE. \url{https://doi.org/10.1117/12.2511516}}
\end{textblock*}

The extreme peak intensities of ultrashort laser pulses lead to linear and non-linear interaction processes with almost all conceivable materials\cite{Feng1997, Nolte1997, Couairon2007}. These remarkable temporal laser properties in combination with tailored spatial focus distributions enable the development of completely new material processing strategies or the optimization of existing ones. The processing of transparent materials\cite{Gattass2008, Shimotsuma2003, Kumkar2014}  -- as a particularly attractive and challenging example -- has become a topic of late, driven by the need for, e.g., efficient machining of display and cover glasses for consumer electronics with increasingly complex geometries\cite{Mathis2012, Bergner2018b, Jenne2018}. This requires a controlled energy deposition at the surface or inside the processing volume and, thus, beam shaping concepts in all three spatial dimensions\cite{Kumkar2017, Mathis2012, Bergner2018a, Jenne2018}. Although demonstrated in the following for selected examples, the potential for structured light concepts shown here is much greater and could not only be employed for a diversity of ultrafast materials processing strategies but also for applications based on continuous-wave lasers, working in spatial single- or multi-mode regime.

In this paper we present three beam shaping concepts and their implementation into industrial materials processing strategies. Starting with the versatile digital-holographic generation of Bessel-like beams exhibiting adapted transverse and longitudinal focus properties we apply sensitive aberration-correction for full thickness, single pass cleaving of glass with thicknesses of up to $\unit[10]{mm}$. Here, vertical glass edges are demonstrated as well as edges of arbitrary shape.

In order to completely utilize the available high average powers (several $\unit[100]{W}$) and pulse energies (several $\unit[1]{mJ}$) of today's industrial ultrafast laser sources\cite{AMPHOSAMPHOS} and, thus, to offer attractive industrial applications, diffractive 3D-beam splitting approaches are presented as second beam shaping concept. This allows to arbitrarily distribute a high number of single foci with individual shape and power at the surface or within the volume of the workpiece and, for example, to increase melting rates for glass welding applications.

Finally, we discuss the digital-holographic generation of flat-top beam profiles for ultrafast micron-scaled ablation, drilling and marking of solids and present processing results of high-spatial frequency masks in thin metal sheets with tailored drilling geometries and taper angles.

For processing of transparent materials, advanced spatial and temporal in situ diagnostics is employed to unveil the fundamentals of light-matter interaction and, in combination with flexible digital-holographic beam shaping techniques, to find the optimal beam shape for the respective laser application. Finally, the high quality of presented beam shaping concepts is proven by implementing the optical concepts into successful laser machining processes. In short, the present work demonstrates the enormous benefits of using structured light concepts\cite{Rubinsztein-Dunlop2016} for industrial ultrafast laser application processes.

We would like to address at this point that pulse durations of employed ultrafast laser sources (TruMicro series 2000 and 5000) are always greater than $\unit[300]{fs}$ (resulting spectral width $\Delta\lambda \lessapprox \unit[5]{nm}$ at $\lambda = \unit[1030]{nm}$) and therefore concepts for the compensation of chromatic aberrations are not necessary, yet (at the applied spatial frequencies). Owing to these (comparatively) long pulse durations, we will also not yet focus on simultaneous spatial and temporal focusing concepts (SSTF)\cite{Zhu2005, Kammel2014}. \\
The paper is organized as follows. In \hyperref[sec:mainBESS]{Sec.\,\ref{sec:mainBESS}} theoretical concepts are introduced and ultrafast processing experiments are conducted based on the use of Bessel-like beams. The realization and potential applications of ultrafast 3D-focus distributions are discussed in \hyperref[sec:main3D]{Sec.\,\ref{sec:main3D}}. Finally, \hyperref[sec:SLM]{Sec.\,\ref{sec:SLM}} treats the concepts of flexible digital-holographic marking and drilling. 

\section{Ultrafast Bessel-like Beams}\label{sec:mainBESS}
The class of Gaussian beams represent stable solutions of classical resonators and therefore undoubtedly form the most famous class of spatial laser radiation. Directly behind them, however, there are probably Bessel-like beams. The areas of application for this beam concept are enormously diverse ranging from atom trapping and particle manipulation\cite{Arlt2000, McGloin2003} to different materials processing strategies\cite{Bhuyan2010, Zhang2018}.

\begin{figure}[t]
	\begin{center}
		\includegraphics[width=1.0\columnwidth]{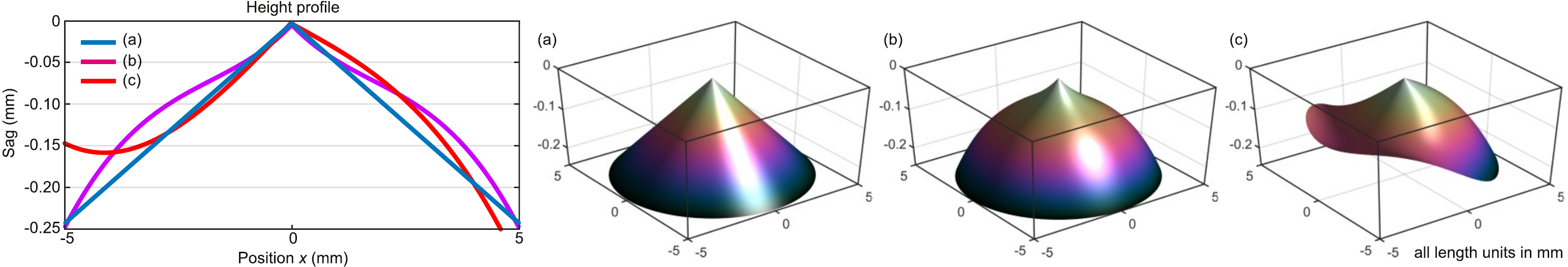}
	\end{center}
\vspace{-0.2cm}
	\caption{Height profiles of an axicon (a) and two axicon-like lenses, (b) and (c). Weakly deforming the ideal axicon, while maintaining radial symmetry (b), allows to manipulate the longitudinal propagation behavior and yields homogenized Bessel-like beams, see \hyperref[fig:mainBESS]{Fig.\,\ref{fig:mainBESS}\,{(e)}}\cite{Flamm2015}. The symmetry of the axicon ``variant'' shown in (c) has been broken completely using a basis of Zernike polynomials. Such free-form axicon lenses can be used to compensate\cite{Jenne2018} or to control\cite{Dudutis2018} wavefront aberrations, see \hyperref[fig:mainBESS]{Fig.\,\ref{fig:mainBESS}\,{(i)}}.}
	\label{fig:Axicons}
\end{figure}

The first Bessel-like beams were generated by ring-slit apertures (far-field generation)\cite{Durnin1987}. However, for efficiency reasons, the axicon-based generation (near-field)\cite{McGloin2005} is preferred. An axicon is a conically ground lens, see \hyperref[fig:Axicons]{Fig.\,\ref{fig:Axicons}\,(a)} and, by design, completely defined by axicon angle $\gamma$\footnote{The widely spread axicon angle definition is used where $\gamma = \left(180^{\circ}-\alpha\right)/2$, with axicon apex angle $\alpha$, see, e.g., McGloin and Dholakia\cite{McGloin2005}.} and refractive index $n$. 
If illuminated by a plane wave (of constant amplitude and phase) with a sufficient degree of coherence, a special interference pattern will be observed starting directly behind the axicon tip that is characterized by an elongated intensity maximum on the optical axis surrounded by weaker, equally spaced rings -- a so-called Bessel-like beam. The peak intensity ratio of the first ring to the central maximum is about $0.17$ and, thus, particularly favorable for nonlinear absorption processes.\footnote{Usually, this ratio is increased significantly, if  modified Bessel-like beams are considered [cf. \hyperref[fig:mainBESS]{Fig.\,\ref{fig:mainBESS}\,(b)--(i)}].} This interference pattern is a consequence of field components propagating along a cone towards the optical axis with refraction angle $\beta \approx \left(n-1\right)\gamma$ and radial component of the wavevector $k_r = 2\uppi\left(n-1\right)\gamma/\lambda$, respectively (in thin-element approximation)\cite{McGloin2005, Leach2006}. The resulting central core spot size of the fundamental Bessel beam then reads as $r_0 = 2.405/k_r$ \cite{McGloin2005}. Illuminating the axicon with a plane wave of radius $R$ allows to approximate the resulting beam length according to $l_0 = R/\left[\left(n-1\right)\gamma\right]$ \cite{McGloin2005}. With the last two equations it is easy to estimate that the aspect ratio of longitudinal to transverse dimensions $l_0/r_0$ could easily reach factors of higher than $10000$, see, e.g. \hyperref[fig:mainBESS]{Fig.\,\ref{fig:mainBESS}\,(a)} -- one of several outstanding properties of this beam class. Further should only be mentioned here in brief: non-diffracting\cite{Durnin1987}, self-healing\cite{McGloin2005}, simple and efficient generation\cite{McGloin2005} and -- particular interesting for the processing of transparent materials -- a natural resistance to spherical aberrations\cite{Jenne2018}.

Despite noteworthy progress in fabrication technology for aspheres or free-form lenses \cite{AsphericonTechnologies, PRP} allowing to realize examples shown in \hyperref[fig:Axicons]{Fig.\,\ref{fig:Axicons}\,(b)} and \hyperref[fig:Axicons]{(c)} we will focus on the diffractive realization for the following reason. Elements for phase modulation are required which imprint phase jumps or singularities to the illuminating optical field. Corresponding free-form lenses would neither be continuous nor differentiable and, thus, difficult to manufacture. We therefore review the diffractive realization (\hyperref[sec:bessel]{Sec.\,\ref{sec:bessel}}) and present degrees of freedom for arbitrary Bessel-like beam shaping in three spatial dimensions by applying a single-element approach. Further, we proof that outstanding beam properties can actually be used to control the deposition of energy in the volume of the workpiece (\hyperref[sec:bessel2]{Sec.\,\ref{sec:bessel2}}) and discuss glass cutting processing results (\hyperref[sec:bessel3]{Sec.\,\ref{sec:bessel3}}). Finally, the implementation of the optical concepts into an industrial cleaving optics is demonstrated (\hyperref[sec:bessel3]{Sec.\,\ref{sec:bessel4}}).

\subsection{Diffractive Beam Shaping Concept}\label{sec:bessel}
Various grating concepts are conceivable for the generation of Bessel-like beams. However, for efficiency-reasons we focus on blazed phase-only gratings theoretically reaching diffraction efficiencies of $\unit[100]{\%}$. A radial symmetric transmission according to
\begin{equation}\label{eq:1}
T^{\text{ax}}\left(r\right) = \exp\left(\imath k_r r\right)
\end{equation}
acts as axicon hologram\cite{Leach2006} [see also \hyperref[fig:phama]{Fig.\,\ref{fig:phama}\,(a)}] with the above defined radial component of the wavevector $k_r$. By this definition a refractive axicon can be assigned directly to its holographic counterpart.\footnote{A discussion on fundamental differences between diffractive and refractive axicons is provided by Leach \textit{et al.}\cite{Leach2006}.} \\
Refractive or reflective axicons have already been demonstrated with axicon angles of $\gamma > 30^{\circ}$ \cite{Boucher2018}. The diffractive realization of such high axicon angles, however, would be associated with enormous fabrication efforts since $\gamma = 20^{\circ}$ already results in grating periods of $p_r = 2\uppi/k_r \approx \unit[5]{\upmu m}$ at $\lambda = \unit[1]{\upmu m}$ which would require grey-tone electron-beam lithography for high diffraction efficiencies. By means of less expensive grey-tone laser-beam lithographic techniques $\gamma$-values of $10^{\circ}$ ($p_r \approx \unit[10]{\upmu m}$ at $\lambda = \unit[1]{\upmu m}$) can be reached. Liquid-crystal-on-silicon based spatial light modulators (SLMs) with pixels of pitch $\approx \unit[10]{\upmu m}$ are able to display digital axicons with angles up to $\approx 3^{\circ}$. However, these axicons with low $\gamma$-values connected to the diffractive concepts can be increased virtually using simple telescopic setups\cite{Brzobohaty2008, Kumkar2017a} limited by the available NA of microscope objectives, see also \hyperref[sec:bessel3]{Secs.\,\ref{sec:bessel3}} and \hyperref[sec:bessel4]{\ref{sec:bessel4}}.

\begin{figure}[t]
	\begin{center}
		\includegraphics[width=0.8\columnwidth]{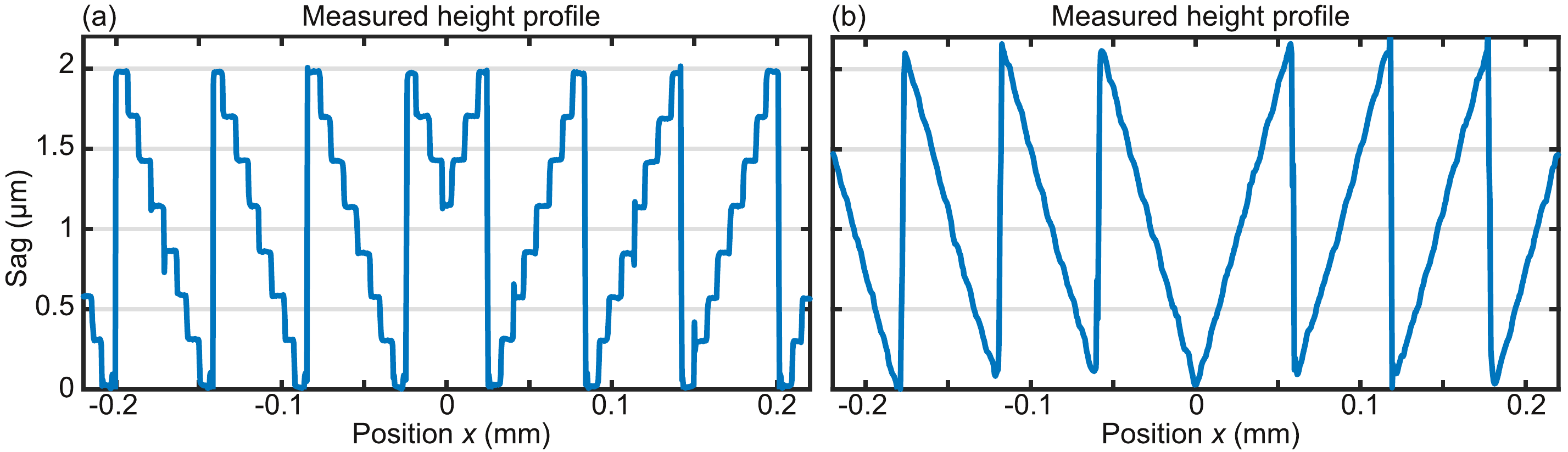}
	\end{center}
\vspace{-0.2cm}
	\caption{Measured height profiles (central details) of diffractive axicons using laser-scanning microscopy. 8-level DOE with first-order diffraction efficiency of $>\unit[90]{\%}$ and zero-order power of $<\unit[0.3]{\%}$ (a). Fresnel axicon-type DOE fabricated via grey-tone lithography with first-order diffraction efficiency of $>\unit[99]{\%}$ and zero-order power amount $<\unit[0.5]{\%}$ (b).}
	\label{fig:surfass2}
\end{figure}

In the following experiments various Bessel-like beams have been generated using either flexible SLMs or diffractive optical elements (DOEs) fabricated via lithography. The desired phase modulation $\Delta \Phi = k_0\Delta L$ is realized by spatially setting optical path differences $\Delta L\left(x,y\right)$. In case of using SLMs this is done electro-optically by controlling the birefringence of the liquid crystals and, thus, $\Delta L\left(x,y\right) = \Delta n\left(x,y\right) H$ and in case of using DOEs, on the other hand, by an etched height profile  $\Delta L\left(x,y\right) =  n \Delta H\left(x,y\right)$ in fused silica. The measured height profile of two diffractive axicons with stated efficiencies are depicted in \hyperref[fig:surfass2]{Fig.\,\ref{fig:surfass2}}.

\begin{figure}[t]
	\begin{center}
		\includegraphics[width=1.0\columnwidth]{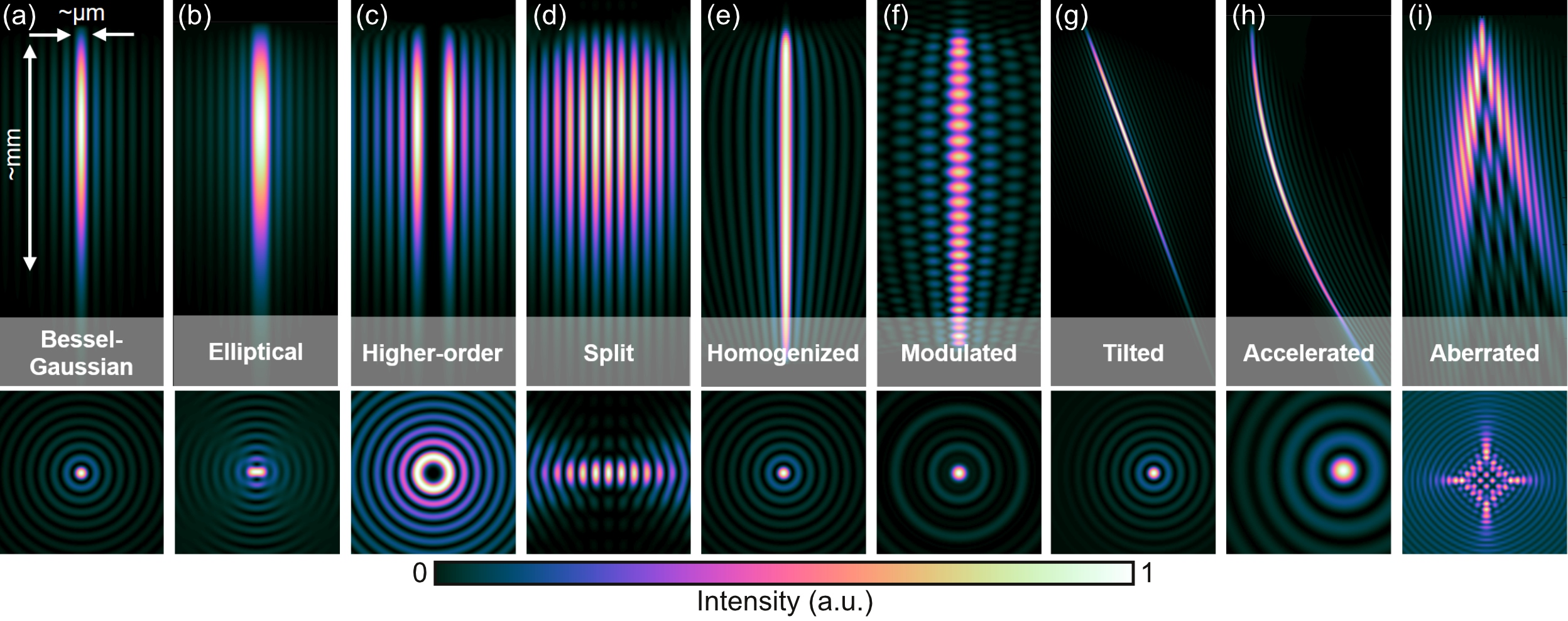}
	\end{center}
\vspace{-0.2cm}
	\caption{Selection of Bessel-like beams used for materials processing (a)-(i) \cite{Flamm2015, Flamm2017, Meyer2017, Du2014, Chremmos2012, Jenne2018}. All cases show the propagation behavior (cross section, top) and corresponding transverse intensity profile at a selected propagation distance (bottom). 
	Here and in some of the following figures, we make use of David Green's color scheme representation.\cite{Green2011}}
	\vspace{0.35cm}
	\label{fig:mainBESS}
\end{figure}

Apart from such advantages as flexibility\cite{Bergner2018a} or the avoidance of round axicon tips\cite{Brzobohaty2008}, diffractive and/or digital holographic techniques are characterized in particular by a large number of degrees of freedom in tailoring the three-dimensional propagation behavior of Bessel-like beams. This plethora of beam shaping possibilities is demonstrated in \hyperref[fig:mainBESS]{Fig.\,\ref{fig:mainBESS}} (without claiming completeness) where the propagation behavior and corresponding intensity cross section of the fundamental Bessel-Gaussian beam [\hyperref[fig:mainBESS]{(a)}] can be compared to several Bessel-like solutions exhibiting adapted intensity distributions in longitudinal and transverse direction [\hyperref[fig:mainBESS]{(b)--(i)}]. 
Simple modifications of the axicon-generating phase mask $T^{\text{ax}}$ [cf. \hyperref[fig:phama]{Fig.\,\ref{fig:phama}}] allow to realize elongated focus distributions with elliptical central spots\cite{Kumkar2017b} [\hyperref[fig:mainBESS]{(b)}] as well as parallel-running split beams\cite{Meyer2017} [\hyperref[fig:mainBESS]{(d)}], Bessel-like beams exhibiting longitudinal homogenization\cite{Flamm2015} [\hyperref[fig:mainBESS]{(e)}] or modulation\cite{Du2014} [\hyperref[fig:mainBESS]{(f)}], and propagating along tilted [\hyperref[fig:mainBESS]{(g)}]\cite{Jenne2018a} or accelerating trajectories [\hyperref[fig:mainBESS]{(h)}]\cite{Chremmos2012}.

\subsection{Bessel-like Beams of Higher-order} \label{sec:bessel2}
\begin{figure}[t]
	\begin{center}
		\includegraphics[width=0.9\columnwidth]{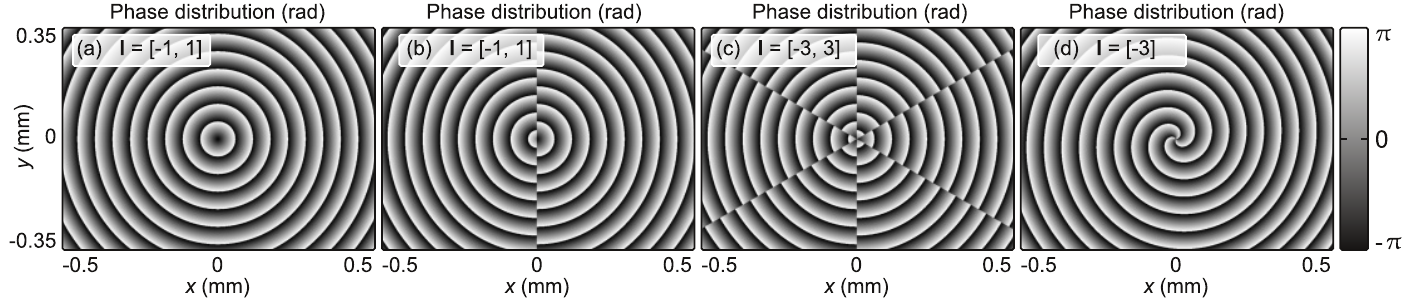}
	\end{center}
\vspace{-0.2cm}
	\caption{Central details of phase distributions $\arg{\left[T\left(x,y\right)\right]}$ for generating Bessel-like beams of different order\cite{Dudley2013, Bergner2018a}. Zero order $\mathbf{l} = \ell_1 =0$ (a), first-order, petal-like, $\mathbf{l} = \left[-1, 1\right]$ (b), third order, petal-like $\mathbf{l} = \left[-3, 3\right]$ (c), as well as the pure Bessel-like beam of third order $\mathbf{l} = \ell_1 = -3$ (d). }
	\label{fig:phama}
\end{figure}

In the following we discuss in detail the diffractive generation of Bessel-like beams of higher-order [cf. \hyperref[fig:mainBESS]{Fig.\,\ref{fig:mainBESS}\,(c)}]. One way to realize such is to start with the axicon transmission function of \hyperref[eq:1]{Eq.\,(\ref{eq:1})} and to multiplex superpositions of azimuthal phase components
\begin{equation}
T^{\text{az}}\left(\phi\right)=\exp{\left\{\imath\arg{\left[\sum_{j}\exp\left(\imath \ell_j \phi\right)\right]}\right\}}, \ell_j \in \mathbb{Z}. 
\end{equation}
The resulting transmission functions for generating Bessel-like beams of different order with corresponding hellicity indices $\mathbf{l} = \left[\ell_1, ..., \ell_j\right]$ are depicted in \hyperref[fig:phama]{Fig.\,\ref{fig:phama}}.
Although the transverse intensity profile of higher-order Bessel-like beams completely differ from the zero order version, the above mentioned remarkable properties of this beam class remain intact. Ring profiles with controllable diameters [c.f. \hyperref[fig:mainBESS]{Fig.\,\ref{fig:mainBESS}\,(c)}] are as possible as the generation of petal beams and those with a certain preferred direction\cite{Bergner2018a, Trichili2014, Dudley2013}. Such intensity distributions/beam shaping approaches show promising prospects for material modifications and processing, e.g., the precise cutting of glass \cite{Bergner2018a, Xie2015,Kumkar2017b} by making targeted use of crack formations\cite{Meyer2017, Jenne2019}.

We verify excellent suitability of this class of beams for transparent materials processing using a time resolved tomographic imaging concept, see Bergner\,\textit{et\,al.}\cite{Bergner2018a}, which allows to reconstruct the three-dimensional spatial distribution of the transient extinction coefficient $\kappa\left(\mathbf{r}\right)$.\footnote{Typically, using transverse pump-probe microscopy there is access to the optical depth $\tau = \ln {\left(I_0/I_S\right)}$ from recording shadowgraph images $I_\text{S}$ and corresponding background signals $I_0$\cite{Grossmann2016}. Considering Lambert-Beer's law, the optical depth represents the integral over the local extinction coefficient $\kappa\left(\mathbf{r}\right)=\kappa\left(x,y,z\right)$ along a certain direction, e.g., $y$-axis: $\tau\left(x, z\right) = \int{\text{d}y \, \kappa\left(x,y,z\right)}$\cite{Grossmann2016, Bergner2018a}.}
\begin{figure}[t]
	\begin{center}
		\includegraphics[width=0.9\columnwidth]{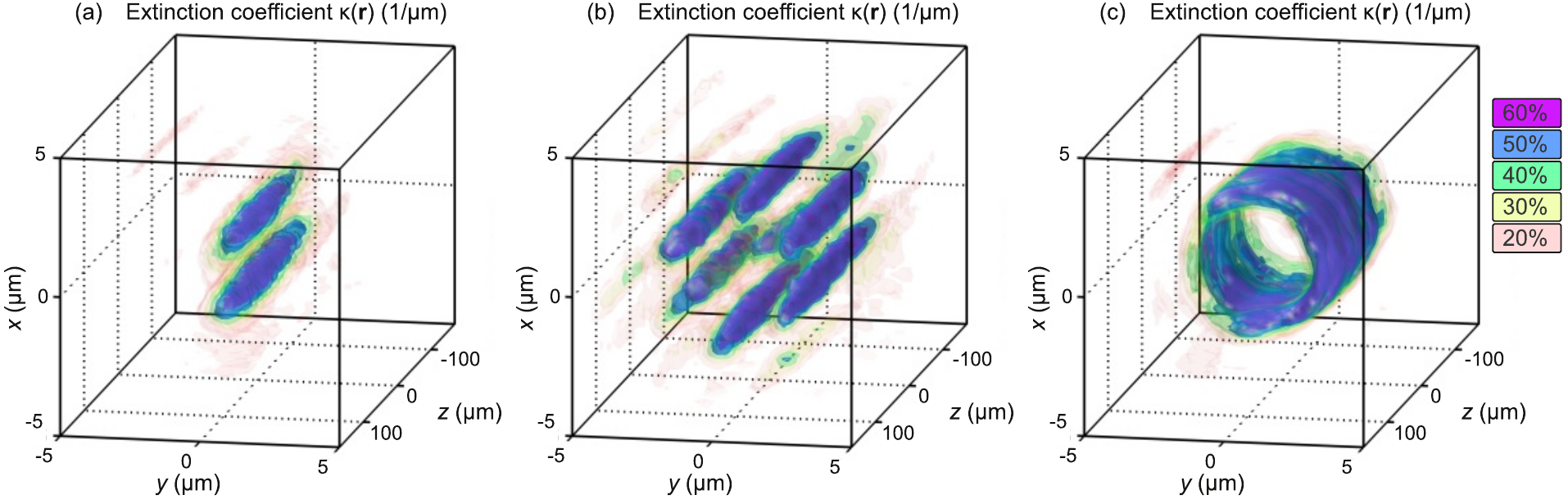}
	\end{center}
\vspace{-0.2cm}
	\caption{Isosurface representation of the reconstructed extinction distribution $\kappa\left(\mathbf{r}\right)$ caused by focusing the Bessel-Gaussian beam of first order, petal-like (a), third-order, petal-like (b) and pure third-order (c) into the sample. Recordings were taken  after a delay of $\unit[7.5]{ps}$. The colored surfaces represent five extinctions thresholds relative to $\kappa_{\text{max}} = \unit[1.2]{\upmu m^{-1}}$ (a), $\kappa_{\text{max}} = \unit[0.4]{\upmu m^{-1}}$ (b) and $\kappa_{\text{max}} =\unit[0.1]{\upmu m^{-1}}$ (c), respectively. Please note the reconstruction volume of $\unit[\left(10 \times 10 \times 300\right)]{\upmu m^3}$, $\left(x \times y \times z\right)$ and the stretched axis scaling in $z$-direction, respectively.}
	\label{fig:extinction}
\end{figure}
This enables to directly measure the material's absorbing response\footnote{The interaction between ultrashort pulses and matter is a notoriously complex process that will not be further explained in this work, please see, e.g., Itoh \textit{et al.}\cite{Itoh2006} or Rethfeld \textit{et al.} \cite{Rethfeld2004} and references therein.} caused by the ultrashort pulse and, thus, to analyze where energy was deposited. Results of this investigation are shown in the isosurface representation of \hyperref[fig:extinction]{Fig.\,\ref{fig:extinction}} where the extinction coefficient $\kappa\left(\mathbf{r}\right)$ is reconstructed after focusing higher-order Bessel-Gaussian beams into Gorilla glass\cite{Bergner2018a}. \hyperref[fig:extinction]{Figure \ref{fig:extinction}\,(a)} shows an elongated $\kappa$-distribution of a few hundred $\unit[]{\upmu m}$ length exhibiting two distinct maxima of distance $\unit[2]{\upmu m}$. In the second case [\hyperref[fig:extinction]{(b)}], the reconstructed extinction distribution reveals an elongated behavior with six parallel running petals distributed point-symmetrically around the optical axis. Finally, \hyperref[fig:extinction]{Fig.\,\ref{fig:extinction}\,(c)}, shows an absorption zone of hollow cylindrical symmetry exhibiting $\approx \unit[3]{\upmu m}$ diameter and a few hundred $\unit[]{\upmu m}$ length. We conclude that the spatial distribution of the \textit{non-linear} absorbing response strongly resembles the \textit{linear} intensity distribution of the corresponding higher-order Bessel-like beam [first-order, petal-like in \hyperref[fig:extinction]{(a)}, third-order, petal like \hyperref[fig:extinction]{(b)} and third-order, pure in \hyperref[fig:extinction]{(c)}]. The reason for this can be found in the remarkable self-healing properties of Bessel-like beams which holds for their higher-order versions as well.\footnote{Please note, that the absorption behavior will be completely different if ultrashort Gaussian beams are focused into transparent materials, due to the lack of self-healing, see Grossmann \textit{et al.}\cite{Grossmann2016} or \hyperref[sec:3D2]{Sec.\,\ref{sec:3D2}}, respectively.}

\subsection{Glass Cutting Using Aberration-corrected Bessel-like Beams}\label{sec:bessel3}
The usage of diffractive optical concepts for the correction of potentially aberrated Bessel-like beams is discussed in the following. In particular, there are aberrations of interest that may occur when propagating through optical interfaces, such as tilted or cylindrical glass surfaces, to cleave transparent materials, e.g., with tailored edges \cite{Jenne2018}. Resulting phase distortions usually yield a characteristic interference pattern with high intensities no longer confined to the optical axis, as can be seen in \hyperref[fig:mainBESS]{Fig.\,\ref{fig:mainBESS}\,(i)}, accompanied by a tremendous loss of peak intensity, prohibiting a successful material modification process. It is our aim to precompensate phase deviations by diffractive or digital holographic approaches and to restore the original Bessel-like focus distribution [cf. \hyperref[fig:mainBESS]{Fig.\,\ref{fig:mainBESS}\,(a)}].

Our design concept makes use of the aforementioned diffractive Bessel-like beam generation (\hyperref[sec:bessel]{Sec.\,\ref{sec:bessel}}) in combination with a $4f$-setup (telescopic demagnification of $M=20$). Thus, the axicon is placed virtually onto the optical interface -- now with increased axicon angle $\gamma' = M\gamma$. As an example we discuss in the following the situation behind tilted plane interfaces\cite{Jenne2018}. However, the procedure is straight forward for arbitrarily shaped surfaces.

\begin{figure}[t]
	\begin{center}
		\includegraphics[width=1.\columnwidth]{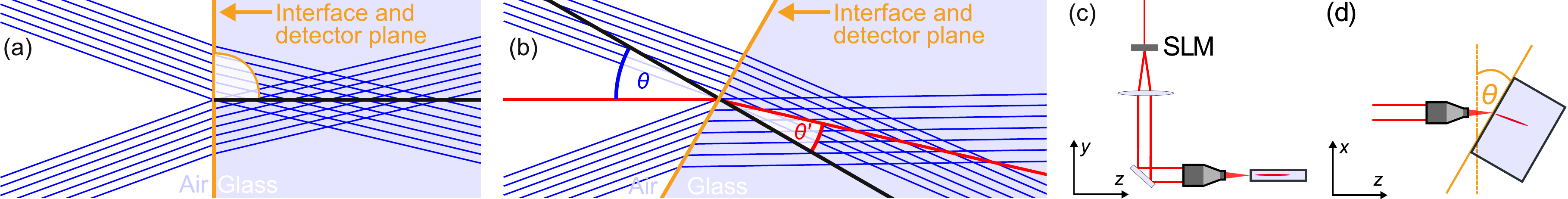}
	\end{center}
\vspace{-0.2cm}
	\caption{Geometric-optical representation of the Bessel-like focus situation behind the glass surface for incident illumination $\theta = 0$ (a) and for inclined illumination $\theta > 0$ (b). Corresponding schematic of the transverse pump-probe microscope (c). Pulses at $\unit[1030]{nm}$ and $\unit[900]{fs}$ pulse duration are beam-shaped by the SLM and focused into the sample using a $4f$-setup. Rotatable mounted glass with definition of cleaving angle $\theta$ (d).}
	\vspace{0.35cm}
	\label{fig:tiler}
\end{figure}
\hyperref[fig:tiler]{Figure \ref{fig:tiler}} shows a geometric-optical representation of an incident Bessel-like beam for perpendicular [\hyperref[fig:tiler]{(a)}] and tilted [\hyperref[fig:tiler]{(b)}] illumination. For the first case, at the interface between the two media, all radial field components exhibit the same absolute angle of incidence value and propagate identical optical path lengths until they interfere constructively to the desired elongated focus. Considering the scenario of a tilted glass surface, the spatially varying angles of refraction determined by Snell's law will break the radial symmetry and, depending on the tilt angle, will result in an interference pattern along the geometric focus zone [\hyperref[fig:mainBESS]{Fig.\,\ref{fig:mainBESS}\,(i)}].\cite{Jenne2018}

\begin{figure}[t]
	\begin{center}
		\includegraphics[width=1.\columnwidth]{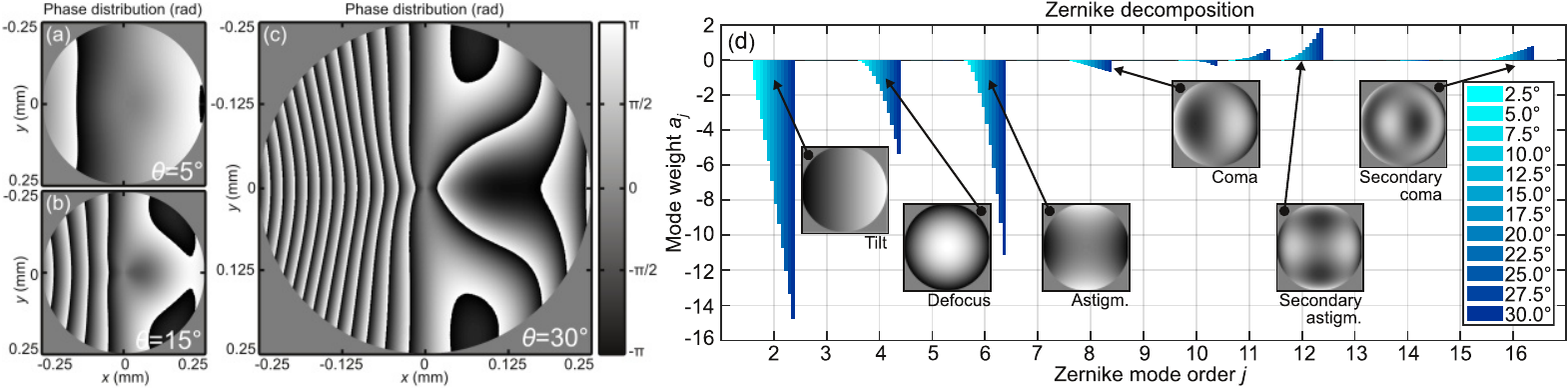}
	\end{center}
\vspace{-0.2cm}
	\caption{Simulated phase aberrations caused by the beam transition through the glass interface for tilt angles of $\theta = \unit[5]{^\circ}$ (a),  $\theta = \unit[15]{^\circ}$ (b) and  $\theta = 30^\circ$ (c). Decomposition of $\Delta\phi\left(x,y\right)$ -- some of them depicted in (a)--(c) -- into a set of $15$ Zernike modes \cite{Noll1976}. Contributing modes show a continuous growth with increasing tilt angle $\theta$ of the plane glass interface.}
	\label{fig:zernBAR2}
\end{figure}

For further analyzing the problem at hand more quantitatively, we use VirtualLab Fusion \cite{Wyrowski2015} providing access to the optical field behind the tilted interface, see \hyperref[fig:tiler]{Fig.\,\ref{fig:tiler}\,(c)} and \hyperref[fig:tiler]{\,(d)}. For this purpose, field-detectors are virtually placed in the plane of the glass surface for normal incidence, see \hyperref[fig:tiler]{Fig.\,\ref{fig:tiler}\,(a)}, and in the plane perpendicular to the refraction angle for tilted incidence, see \hyperref[fig:tiler]{(b)}, respectively. Knowledge about the aberrated optical field $E^{\text{ab}}$  in the plane of the interface $E^{\text{ab}}\left(x, y\right) = A^{\text{ab}}\left(x, y\right)\exp\left[\imath\phi^{\text{ab}}\left(x, y\right)\right]$ enables to calculate deviations from the undisturbed field $E^{\text{id}}\left(x, y\right) = A^{\text{id}}\left(x, y\right)\exp\left[\imath\phi^{\text{id}}\left(x, y\right)\right]$ generating the ideal Bessel-like beam. Due to small differences for the respective Fresnel transmission coefficients we assume $A^{\text{ab}} \approx A^{\text{id}}$ and focus on deviations of the corresponding phase distributions $\Delta\phi = \phi^{\text{id}} - \phi^{\text{ab}}$. Results of these simulations are depicted in \hyperref[fig:zernBAR2]{Fig.\,\ref{fig:zernBAR2}\,(a)--(c)}, where $\Delta\phi$ is plotted as an example for three different tilt angles $\theta$ defined within a circle of radius $R_{\phi}' = \unit[250]{\upmu m}$.
As expected, phase aberrations are growing for increasing values of $\theta$ [\hyperref[fig:tiler]{Fig.\,\ref{fig:tiler}\,(a)--(c)}]. This is illustrated more clearly in \hyperref[fig:zernBAR2]{Fig.\,\ref{fig:zernBAR2}\,(d)} where we perform a decomposition into a set of Zernike modes $\left\{Z_{j}\left(x,y\right)\right\}$ of order $j$ (normalized and indexed according to Noll \cite{Noll1976}) using $\Delta\phi = \arg\left[\exp\left(\imath\uppi \sum_{j=1}^{j_{\text{max}}} a_jZ_j \right)\right]$ \cite{Schulze2013}. The graph shows weights of coefficients $a_j$ for $12$ different tilt angles. Characteristic for the problem at hand is the continuous growth of certain coefficients with increasing $\theta$ such as $j=4$ (defocus), $6$ (oblique astigmatism) or $12$ (vertical secondary astigmatism).\cite{Jenne2018} \\
As already described earlier, our processing optic design consists of the SLM or DOE, respectively, and a $4f$-setup, see \hyperref[fig:tiler]{Fig.\,\ref{fig:tiler}\,(c)}. This arrangement allows for both Bessel-like beam generation and compensation of occurring aberrations using a single digital hologram or DOE. For this purpose, the calculated phase aberrations $\Delta\phi\left(x,y\right)$ are inverted, the spatial scaling is adapted to the magnification and the resulting transmission is multiplexed with the axicon hologram $T_{\text{ax}}$. Then, the final phase-only hologram $T^{\text{tot}}$ reads as $T^{\text{tot}} = T^{\text{ax}}T^{\text{ab}} = \exp{\left[\imath\left(\beta r - \Delta\phi\right)\right]}$, now defined on a magnified circle of radius $R_{\phi}=MR_{\phi}' = \unit[5]{mm}$, for our particular case.\cite{Jenne2018}

\begin{figure}[t]
	\begin{center}
		\includegraphics[width=1.\columnwidth]{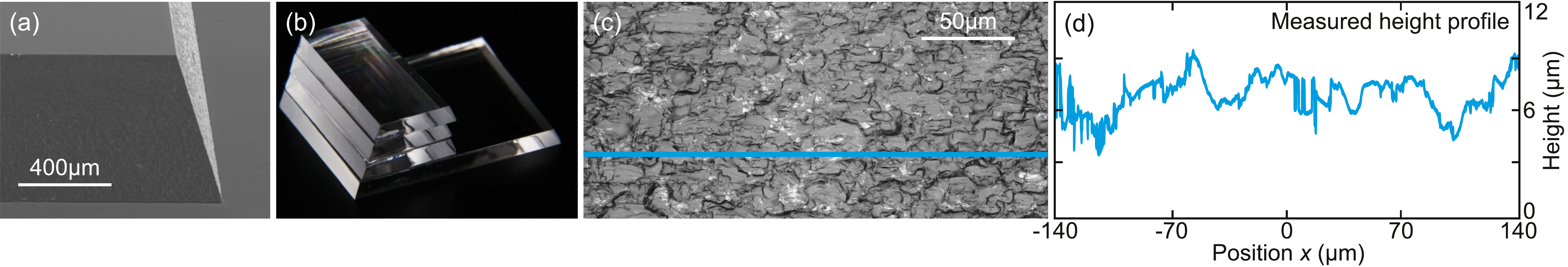}
	\end{center}
	\vspace{-.3cm}
	\caption{Scanning electron microscope image (a) for the cleaved edges with $\unit[30]{^\circ}$ tilting angle in air of $\unit[1]{mm}$ glass and cleaved samples of $\unit[2]{mm}$ thickness in (b), both SCHOTT borofloat 33. Detail of the cleaved surface and its corresponding surface-roughness measurement in (c), (d).\cite{Jenne2018}}
	\label{fig:Cleaving}
\end{figure}

To demonstrate the applicability of our approach, we integrated the resurrected features of the corrected Bessel-like beams into a cleaving process. The employed processing laser was a $\unit[120]{W}$ disk-laser (TruMicro Series 5000) in order to process glasses with thicknesses of $>\unit[2]{mm}$ (SCHOTT borofloat$^{\text{\textregistered}}$ 33). Here we report on a single pass edge cleaving laser application with $\theta'=\unit[20]{^\circ}$ inside the material, cf. \hyperref[fig:Cleaving]{Fig.\,\ref{fig:Cleaving}\,(a),(b)}. We use a spatial pulse distance of $\unit[6]{\upmu m}$ at a processing speed of $\unit[40]{mm/s}$ combined with a pulse train configuration (burst) consisting of four pulses with a temporal delay of $\unit[17]{ns}$ between each consecutive pulses, pulse duration of $\unit[1]{ps}$ and a pulse train energy of approximately $\unit[800]{\upmu J}$. \hyperref[fig:Cleaving]{Figure\,\ref{fig:Cleaving}(a)} shows a scanning electron microscope image of two cleaved glass edges and reveals the precise and defined edge quality for all inclined edges on the $\unit[1]{mm}$ thick glass sample. One exemplary image of the surface quality created with a laser scanning microscope and one corresponding line plot is depicted in \hyperref[fig:Cleaving]{Fig.\,\ref{fig:Cleaving}\,(c)} and \hyperref[fig:Cleaving]{(d)}, respectively. The arithmetic mean roughness for this process is $R_a \approx \unit[0.8]{\upmu m}$ and absolute vertical height offset $R_q \approx \unit[5]{\upmu m}$.\cite{Jenne2018}

\subsection{Implementation into an Industrial Cleaving Optics}\label{sec:bessel4}
The optical concepts discussed in the previous sections have been incorporated into the development of an industrial processing optics, known as TOP Cleave cutting optics\cite{TRUMPFTOPC}. In the near future, the second generation will be available with outstanding specifications presented in the following.

The new, modular optical concept in combination with the corresponding TruMicro laser platform\cite{Jansen2018} allows transparent materials of thicknesses between $\unit[0.3]{mm}$ and $\unit[8]{mm}$ to be modified and cut, respectively. At the same time, the design of this processing optics is characterized by being particularly compact, light and robust and, thus, can be subjected to high accelerations ($\unit[5]{G}$). The modular concept also allows extensions with an axis for adjustment of focus position, a swivel axis or bending mirrors. There will be TOP Cleave versions optimized for use with the TruMicro Series 2000 and 5000 laser sources for infrared ($\lambda = \unit[1030]{nm}$) and green ($\lambda = \unit[515]{nm}$) wavelength to cut colored glasses, such as infrared filter glasses, additionally. The patented optical concept\cite{Kumkar2017a, Kumkar2017b,Kumkar2017c} is based on a diffractive optical element for Bessel-like beam generation and adapted microscope objectives for the mid-NA region of up to $0.4$. Additionally, in designing these objectives, particular emphasis was placed on achieving a long working distance. Finally, it should be noted, that the optic is compatible with all Bessel-like solutions presented in \hyperref[fig:mainBESS]{Fig.\,\ref{fig:mainBESS}}, thus, also special glasses can be processed with customized geometries, with e.g., bevels or tailored edges, cf.\,\hyperref[fig:Cleaving]{Figs.\,\,\ref{fig:mainBESS}\,(g)} and \hyperref[fig:Cleaving]{\ref{fig:Cleaving}}, respectively.
Specifications are summarized in \mbox{\hyperref[tab:table1]{Table \ref{tab:table1}}}. Please note, that dimensions and mass indicated here hold for the smallest possible version which enables material modifications of thicknesses up to $\unit[3]{mm}$.

\begin{figure}[t]
	\begin{center}
		\includegraphics[width=1.\columnwidth]{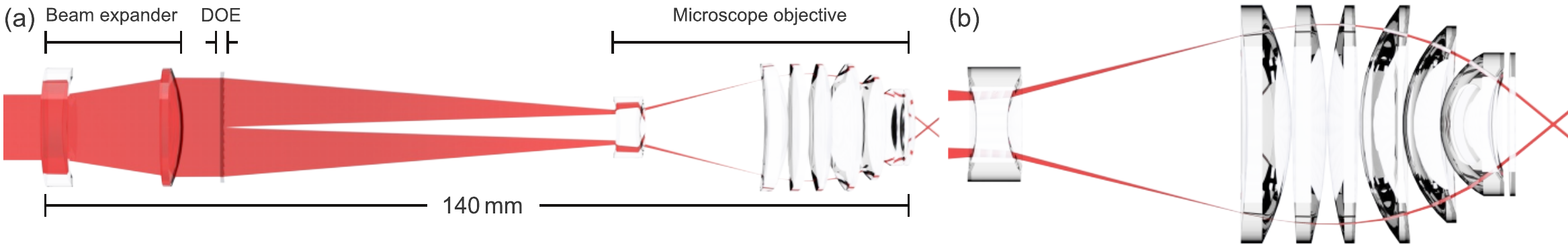}
	\end{center}
\vspace{-0.2cm}
	\caption{Geometric-optical representation of the beam path within TOP Cleave cutting optics (second generation) and focusing of the Bessel-like beam. Beam propagation through beam expander, Bessel-like beam generating DOE and focusing microscope objective (from left to right) (a). Detail of the focusing objective of a high-NA TOP Cleave version ($\text{NA}=0.56$, available upon request) (b). }
	\label{fig:TC2sct}
\end{figure}

The geometric-optical representation of the light path within a \mbox{high-NA} TOP Cleave version is depicted as an example in \hyperref[fig:TC2sct]{Fig.\,\ref{fig:TC2sct}}. After the beam expansion light passes the central beam shaping element generating a virtual Bessel-like beam profile. Using an adapted microscope objective a real Bessel-like beam is formed with demagnification and desired focus length.

\begin{figure}[t]
	\begin{center}
		\includegraphics[width=.75\columnwidth]{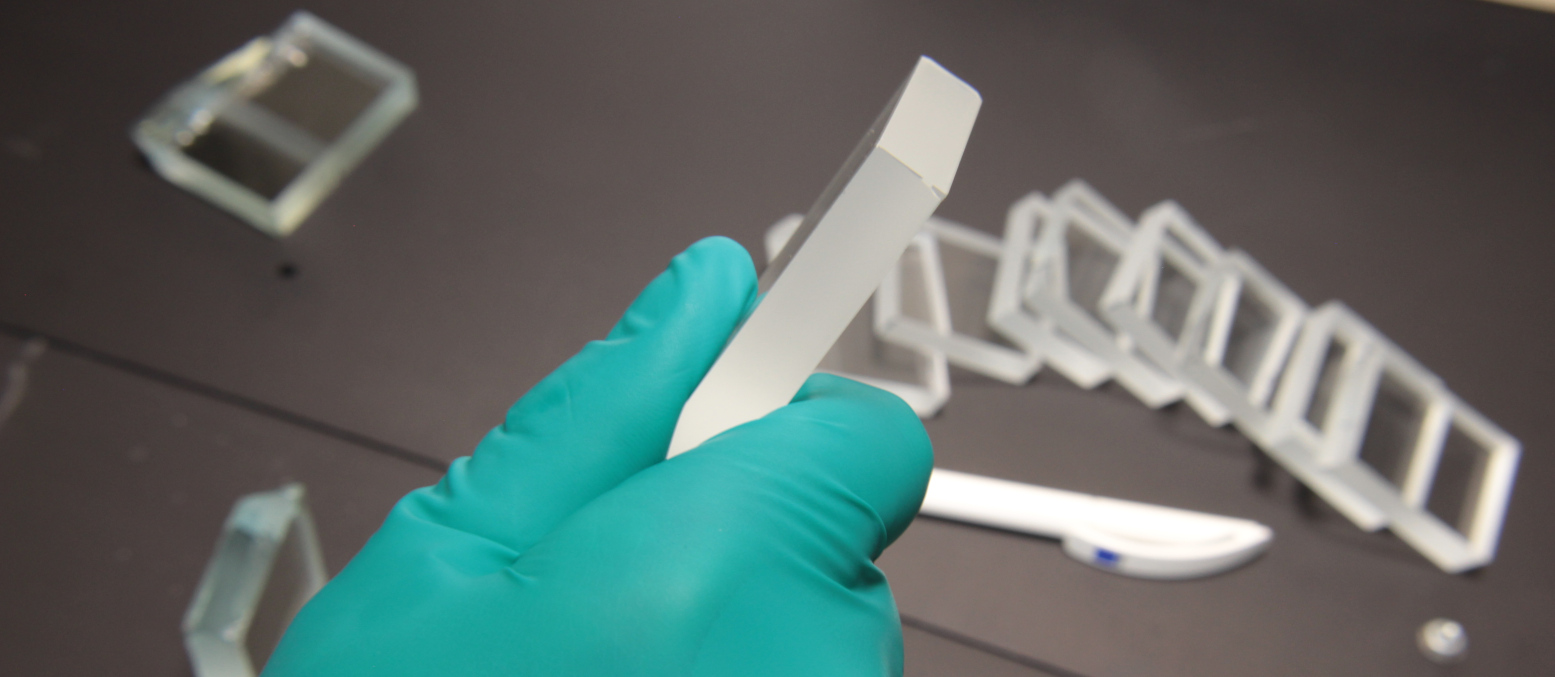}
	\end{center}
\vspace{-0.2cm}
	\caption{Cutting result of $\unit[10]{mm}$ thick soda-lime glass modified using TOP Cleave cutting optics (second generation) and a TruMicro Series 5000 laser source providing pulse energies of $\unit[1.5]{mJ}$ (available in the near future).}
	\label{fig:IMG_0099}
\end{figure}

A remarkable example of a successful glass cutting process can be seen in \hyperref[fig:IMG_0099]{Fig.\,\ref{fig:IMG_0099}}. Here, soda-lime glass of $\unit[10]{mm}$ thickness was modified in a single-pass using TOP Cleave and a TruMicro Series 5000 laser emitting ultrashort pulses of $\unit[1.5]{mJ}$ energy (available in the near future). The glass cutting was achieved by inducing mechanical stress using cut-running pliers. To the best of our knowledge, using a single-pass ultrafast glass modification process no thicker glasses have been cut so far.
\linespread{1.0}
\begin{table}[]
		\centering
		\caption{Technical data of TOP Cleave cutting optics (second generation).}
		\vspace{.1cm} 
		\begin{tabular}{ll}
			\specialrule{.1em}{.1em}{.1em}
			& \vspace{-0.2cm} \\
			\textbf{Structural design\hspace{1cm}} &   \\
			Width & $\unit[40]{mm}$ \\
			Height & $\unit[137]{mm}$ \\
			Depth & $\unit[42]{mm}$ \\
			Minimum movable mass & $\unit[0.29]{kg}$ \vspace{.2cm} \\ 
			\textbf{Laser parameter\hspace{1cm}} &  \\
			Wavelength & $\unit[515]{nm}$ and $\unit[1030]{nm}$  \\
			Laser platform & TruMicro Series 2000 and 5000 \vspace{.2cm} \\
			\textbf{Material properties\hspace{3cm}} &  \\
			Material thicknesses to be modified & $\unit[\left(0.3\,\dots\, 8\right)]{mm}$  \\
			Materials & All common transparent materials, such as\\
			& $\cdot$	Glasses \\
			& \hspace{0.5cm} Fused silica \\
			& \hspace{0.5cm} Soda lime glass \\
			& \hspace{0.5cm} Borosilicate glass \\
			& \hspace{0.5cm} Alumino silicate glass\\
			& \hspace{0.5cm} Alumino borosilicate glass\\
			& $\cdot$ Ceramics \\
			& \hspace{0.5cm} Sapphire \\
			& \hspace{0.5cm} Transparent alumino\\
			& $\cdot$ Glass ceramics \\ \vspace{.2cm}
			\textbf{Optics configuration} \vspace{-.2cm}  &  \\
			Raw beam diameter & $\approx \unit[5]{mm}$   \\
			Focusing microscope objectives & $f = \unit[20]{mm}$, $\text{NA}=0.33$, working distance = $\unit[19.8]{mm}$  \\
			& $f = \unit[10]{mm}$, $\text{NA}=0.42$, working distance = $\unit[7.5]{mm}$ \\
			Beam expansion modules & 0.35, 0.5, 0.75, 1.5, 2.0, 2.8 \vspace{.2cm}\\
			\specialrule{.1em}{.1em}{.1em}  
		\end{tabular}
	\vspace{0.5cm}
		\label{tab:table1}
\end{table}
\linespread{1.0}

\section{Ultrafast 3D-focus Distributions}\label{sec:main3D}
Industrial ultrafast laser sources exhibiting several millijoule of pulse energies and several hundred watts of average power will be available soon\cite{Eidam2010, Fattahi2014, Negel2015, AMPHOSAMPHOS}. This allows the development of completely new application processes, such as, e.g. single-pass, millimeter-scaled cutting of glasses [cf. \hyperref[sec:bessel3]{Secs.\,\ref{sec:bessel3}} and \hyperref[sec:bessel4]{\ref{sec:bessel4}}] or the optimization of existing ones. The potential for optimization is given in particular by increasing throughput through parallel processing and, thus, to exploit the full performance of the laser source.\cite{Kumkar2017} Based on well-known techniques for parallel data recording and storing \cite{Streibl1989,Ren2014, Zhu2014, Gu2014} we use concepts to generate multifocal arrays and extend them to arbitrarily place a multitude of foci within a micrometer-scaled working volume\cite{Jesacher2010, Simmonds2011}. We start with theoretical considerations (\hyperref[sec:3D1]{Sec.\,\ref{sec:3D1}}) and present the successful implementation into laser application processes for ablation and volume modifications, respectively (\hyperref[sec:3D1]{Sec.\,\ref{sec:3D2}}).

\subsection{Diffractive Beam Splitting Concept}\label{sec:3D1}
One way to displace a focus from its original position behind a lens is achieved by introducing phase modifications into the wavefront of the illuminating optical field. In terms of Zernike polynomials\cite{Noll1976}, a proper choice of tip/tilt modes in combination with a defocus allows to control the transverse $\left(\Delta x, \Delta y\right)$ and longitudinal $\Delta z$ displacement, respectively. We extend this simple concept for the manipulation of a single focus to a 3D-beam splitting concept by exploiting the linearity property of optics and multiplex the corresponding holographic transmission functions\cite{Flamm2012, Forbes2016} (one for each focus to be placed in the working volume).

Transverse displacement for the focus of order $j$ is achieved by setting a linear blaze grating with spatial frequency $\mathbf{K}_j = \left( K_{x,j}, K_{x,j}\right)$ in the front focal plane of a lens with focal length $f_{\text{FL}}$. The corresponding transmission function then reads as 
\begin{equation}\label{eq:grat1}
	T_{j}^{\text{grat}}\left(\mathbf{r}\right) = \exp{\left[\imath\left(\mathbf{K}_j\mathbf{r}+\phi_j\right)\right]}
\end{equation}
and yields a transverse displacement according to $\Delta x_j = f_{\text{FL}}\tan{\left[\sin^{-1}{\left(\lambda K_{x,j}/2\uppi\right)}\right]}$.\footnote{Deduced from the grating equation. The procedure is straightforward for the displacement in $y$-direction.} Longitudinal displacement $\Delta z_j$ of the $j$th-order focus is realized by introducing the defocus ``aberration'' using, e.g., a holographic lens transmission with focal length $f_j$
\begin{equation}
T_{j}^{\text{lens}}\left(\mathbf{r}\right) = \exp{\left[\imath\uppi r^2/\left(\lambda f_j\right)\right]}.
\end{equation}
Then, the longitudinal shift is directly deduced from $\Delta z_j \approx -f_{\text{FL}}^2 / \left(f_{\text{FL}}+f_j\right)$.\footnote{In paraxial approximation the following relation holds for the resulting effective focal length $1/f_{\text{eff}} \approx 1/f_{\text{FL}} + 1/f_{j}$.} Combinations of both shifts are achieved by multiplying both transmissions $T_j = T_{j}^{\text{grat}}T_{j}^{\text{lens}}$. Multiplexing these $j_{\text{max}}$ transmission functions will yield the total transmission according to
\begin{equation}\label{eq:tot}
T^{\text{tot}}\left(\mathbf{r}\right) = \sum_j^{j_{\text{max}}} T_{j} = \sum_j^{j_{\text{max}}} T_{j}^{\text{grat}}\left(\mathbf{r}\right)T_{j}^{\text{lens}}\left(\mathbf{r}\right).
\end{equation}
In general, this multiplexing scheme will result in a complex valued transmission $T^{\text{tot}}\left(\mathbf{r}\right) = A^{\text{tot}}\left(\mathbf{r}\right)\exp{\left[\imath\Phi^{\text{tot}}\left(\mathbf{r}\right)\right]}$. Different approaches exist to realize such a transmission as phase-only hologram, see, e.g., Davis \textit{et al.}\cite{Davis1999} or Arriz{\'o}n \textit{et al.}\cite{Arrizon2007}. However, a particularly efficient and simple solution represents $A^{\text{tot}}\left(\mathbf{r}\right) = \mathbb{1}$, thus setting the amplitude modulation to unity and directly use $T^{\text{tot}}\left(\mathbf{r}\right) =\exp{\left[\imath\Phi^{\text{tot}}\left(\mathbf{r}\right)\right]}$ as phase-only transmission. This approach will yield optical powers in unwanted diffraction orders, but, nonetheless, will be significantly more efficient than aforementioned phase-coding techniques. However, it has negative impact on the uniformity of individual spots. To restore equal power distribution a set of constant phase offsets $\left\{\phi_j\right\}$ in the grating representation of \hyperref[eq:grat1]{Eq.\,(\ref{eq:grat1})} can be found by an iterative optimization routine. Here, the optical field in the working volume and the optical power of the $j_{\text{max}}$ spots have to be simulated for each iteration\cite{Leutenegger2006}. This iterative Fourier-transform algorithm \cite{Wyrowski1988} is expanded to all three spatial dimensions (3D--IFTA) and yields the phase offsets $\left\{\phi_j\right\}$ until a desired uniformity is reached. The deduced set of $\left\{\phi_j\right\}$, finally, completely determines the total phase-only transmission $T^{\text{tot}}$ for each spot placed in the working volume by $\left\{\mathbf{K}_j, f_j\right\}$.

\begin{figure}[t]
	\begin{center}
		\includegraphics[width=1.0\columnwidth]{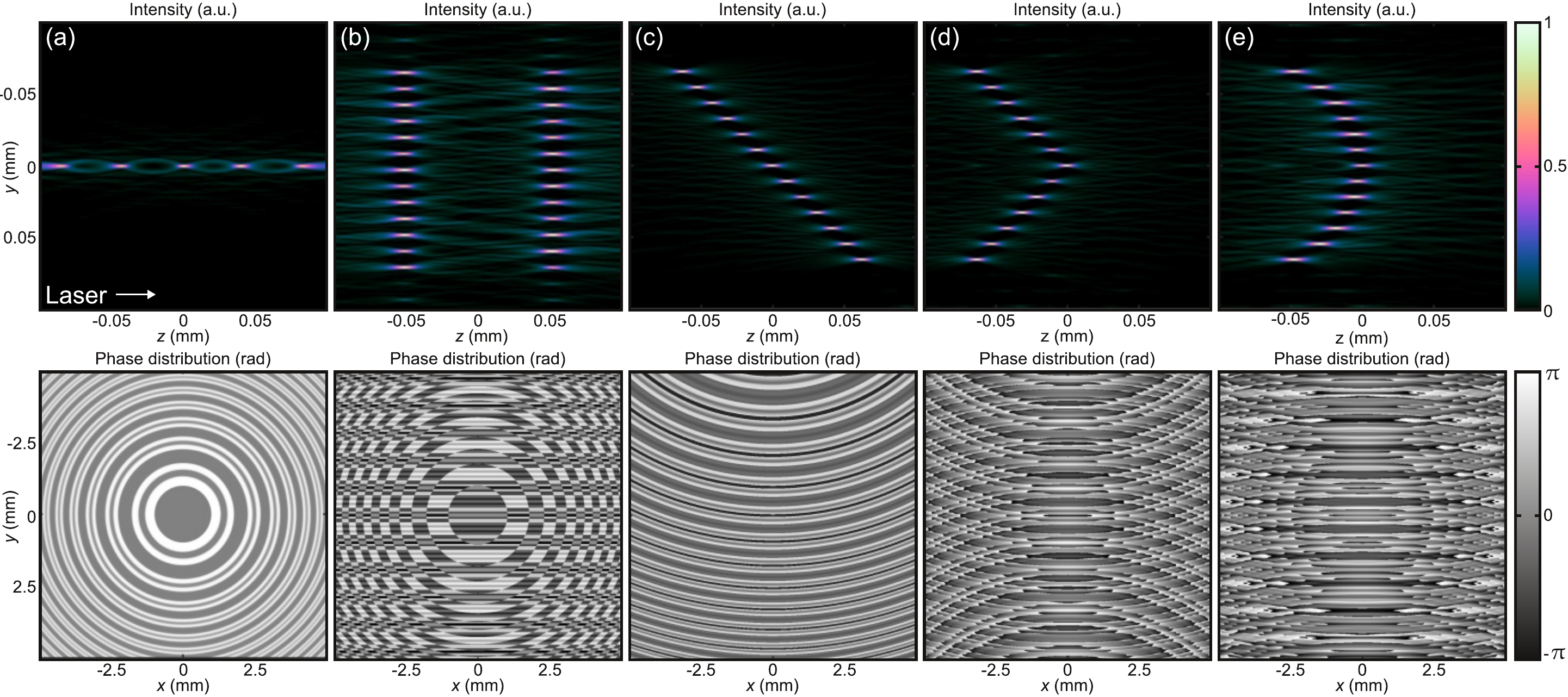}
	\end{center}
\vspace{-0.2cm}
	\caption{Selection of 3D-focus distributions [intensity cross sections, $I\left(x=0, y, z\right)$, top row, $z$-axis corresponds to propagation direction] with corresponding phase-only transmission functions ($\arg{\left[T_{\text{tot}}\left(x,y\right)\right]}$, bottom row) for beam generation. The Gaussian-foci are arbitrarily distributed within the working volume. Regular longitudinal focus arrangement (a). Combination of regular longitudinal and transversal beam splitting (b). Focus arrangement along a $45^{\circ}$-straight line (c). Focus arrangement along a $90^{\circ}$-angle (d). Focus arrangement along an accelerating trajectory (half circle) (e).}
	\label{fig:3dfoc}
\end{figure}

To demonstrate the efficacy of this concept consider the scenarios depicted in \hyperref[fig:3dfoc]{Fig.\,\ref{fig:3dfoc}} where five 3D-focus distributions can be seen with corresponding phase-only transmission functions.
In all simulations a fundamental Gaussian beam propagates from negative to positive $z$-direction, passes the diffractive optical element and is focused in $2f$-configuration with $f=\unit[10]{mm}$. The resulting Gaussian-foci are arbitrarily distributed within the working volume shown as intensity cross section  $I\left(x=0, y, z\right)$. As example a regular longitudinal focus arrangement \hyperref[fig:3dfoc]{Fig.\,\ref{fig:3dfoc}\,(a)}, a combination of regular longitudinal and transversal beam splitting \hyperref[fig:3dfoc]{(b)}, a focus arrangement along a straight line \hyperref[fig:3dfoc]{(c)}, along an angle \hyperref[fig:3dfoc]{(c)}, and a focus arrangement along an accelerating trajectory \hyperref[fig:3dfoc]{(d)} can be seen.

\begin{figure}[t]
	\begin{center}
		\includegraphics[width=1.0\columnwidth]{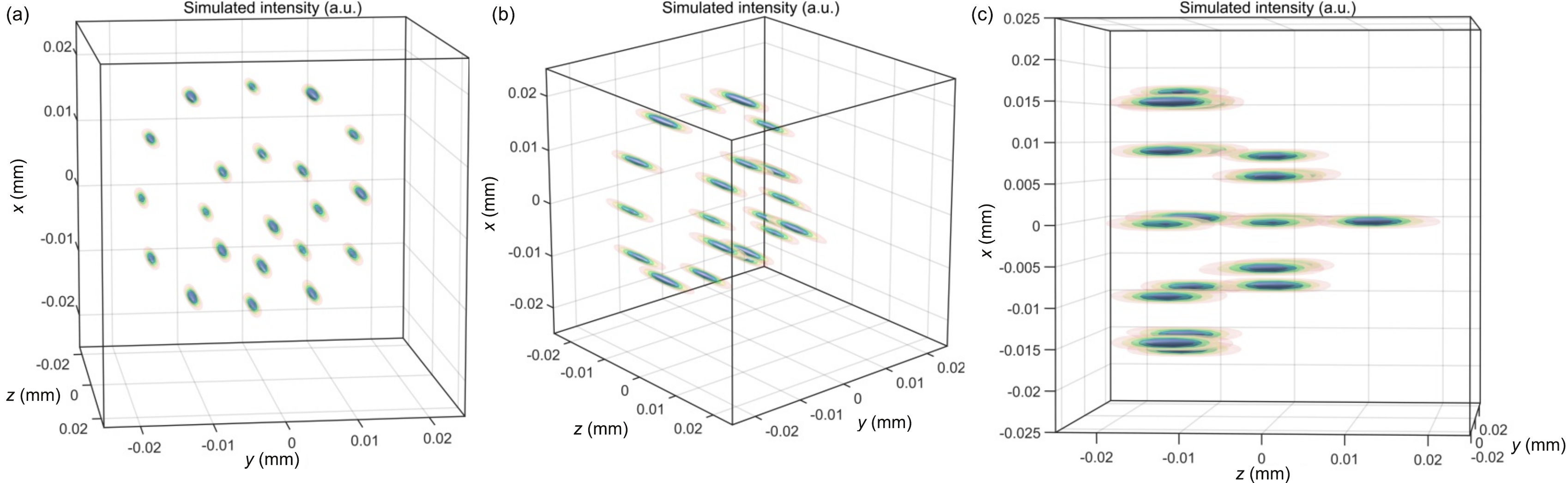}
	\end{center}
\vspace{-0.2cm}
	\caption{Example of a 3D-focus distribution (isosurface representation of intensities) where 21 single foci are distributed along a cone surface within the working volume (cube of edge length $\unit[50]{\upmu m}$). The raw beam of diameter $\unit[5]{mm}$ propagates from negative to the positive $z$-direction, passes the diffractive, beam splitting element and is focused by a $20\times$ microscope objective in $2f$-configuration. All subfigures show the same focus distribution for three different observation directions (a)--(c). Please note axis orientation.}
	\label{fig:3dpure}
\end{figure}

Since the examples of \hyperref[fig:3dfoc]{Fig.\,\ref{fig:3dfoc}}, strictly speaking, show two-dimensional beam splittings only, a further focus distribution is depicted (optical setup identical to previous simulation), see isosurface representation of intensities in \hyperref[fig:3dpure]{Fig.\,\ref{fig:3dpure}}, where 21 single foci are distributed along a cone surface within the working volume (cube of edge length $\unit[50]{\upmu m}$).
The subfigures show the same focus distribution for three different observation directions \hyperref[fig:3dpure]{(a)--(c)}.

\subsection{Material Processing Using Transverse and Longitudinal Beam Splitting}\label{sec:3D2}
The efficacy of the diffractive 3D-beam splitting concept is demonstrated in the following by means of two selected examples. Usually, a measured focus distribution is used as proof for a successful beam shaping concept. In this work, on the other hand, we make use of the processing result and investigate the \textit{impact} of the spatially and temporally shaped laser pulses on the material.

\begin{figure}[t]
	\begin{center}
		\includegraphics[width=1.0\columnwidth]{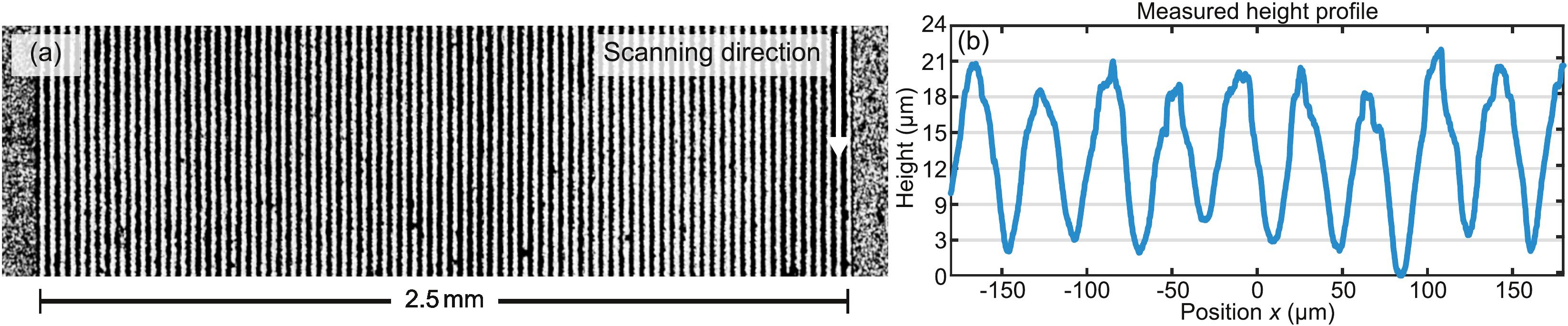}
	\end{center}
\vspace{-0.2cm}
	\caption{Microscope image of groves in steel generated by multi-path ablation using a multi-spot array scanned perpendicular to the beam splitting direction (a) and corresponding height profile measurement of a detail of the processed area (b).}
	\label{fig:bao}
\end{figure}

A first example is chosen to demonstrate the suitability of a multi-spot approach for efficient laser engraving of burr-free grooves in sheet steel with a processing speed of $\unit[>2]{m/s}$. Groove depth and width should exceed $\unit[10]{\upmu m}$. Preliminary examinations on energy specific ablation volume have shown that for $\lambda = \unit[1030]{nm}$ best efficiency was achieved for pulse durations of $\unit[900]{fs}$ at fluences of $\approx \unit[1]{J/cm^2}$.\cite{Kumkar2017} Desired groove dimensions could be reached with multipath engraving of 81 passes at repetition rates of $\unit[800]{kHz}$ using a TruMicro 5070 laser system working in fundamental Gaussian-mode operation $M^2<1.3$. In order to achieve the desired groove in a single pass, a diffractive optical element (DOE) was designed that yields $81$ Gaussian-like spots in a line of $\unit[2.5]{mm}$ length using the employed $f$-$\theta$-objective. The 2-inch DOE was realized as binary height profile etched in fused silica possessing an efficiency of $\unit[82]{\%}$, a uniformity error of $\unit[<5]{\%}$ and a relative zero-order power amount of merely $\unit[<0.1]{\%}$. 
\hyperref[fig:bao]{Figure\,\ref{fig:bao}}
shows a corresponding ablation result with DOE aligned by intention perpendicular to the direction of feed to characterize the quality of the beam shaping concept. Each of the 81 engraved lines was generated by 81 passes (pulse duration $\unit[900]{fs}$, rep.\,rate $\unit[800]{kHz}$, feed rate $\unit[2.2]{m/s}$) and exceed depths of $\unit[10]{\upmu m}$ proving the successful implementation of the beam shaping concept into the scanning system technology. This example clearly demonstrates the suitability of such a multi spot approach for processing along straight paths with high speed. The benefit from using split beams is not solely based on the increased processing speed. No detrimental effects due to heat accumulation are visible at an average power of about $\unit[100]{W}$\cite{Kumkar2017}. We want to emphasize at this point that the high processing speed becomes possible by the laser source and the beam splitting concept but are realized here by the dynamic of the scanning system. In general, large deflection angles ($>\unit[10]{^\circ}$) will result in field curvature aberrations (rotation of the focus line) and in coma/astigmatism aberrations of single spots (loss of peak intensity). Ideally, the design of the beam splitting DOE should take the imaging performance of the scanning objective into account. 

By means of a second example we extend the beam splitting concept to the dimension of the beam propagation direction ($z$-axis).
\begin{figure}[t]
	\begin{center}
		\includegraphics[width=0.6\columnwidth]{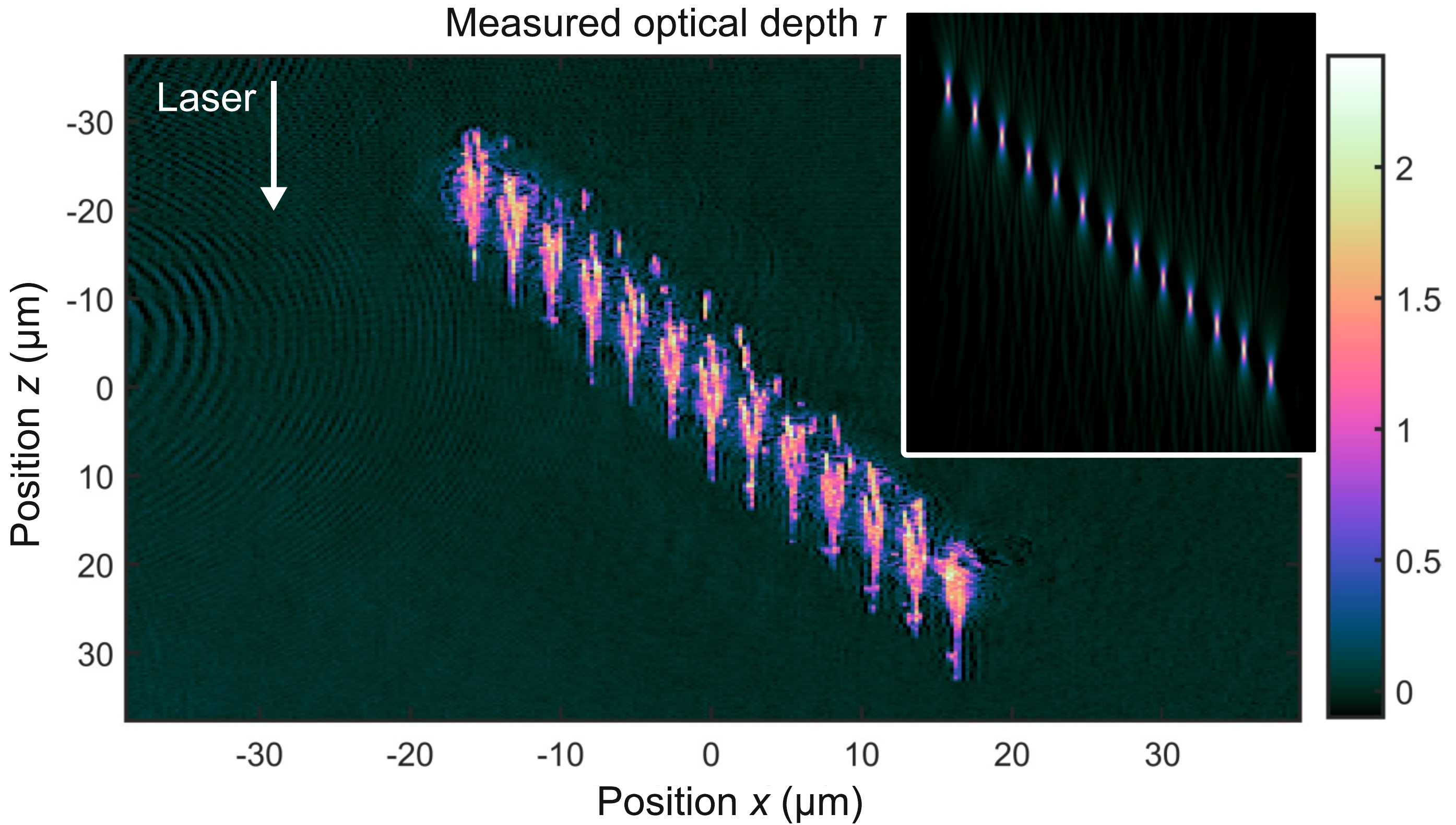}
	\end{center}
\vspace{-0.2cm}
	\caption{Measured optical depth $\tau\left(x, z\right)$ for a multi-spot focus distribution, see inset, inside the volume of borosilicate glass using transverse pump-probe microscopy \cite{Jenne2018b}. The beam propagates from negative to positive $z$-direction. The absorption projection is composed of $13$ single zones and clearly indicates that energy is deposited spatially controlled by the beam splitting concept.}
	\label{fig:scheiss1}
\end{figure}
This enables higher flexibility for distributing the laser pulse energy within the working volume of the material to be processed (here: non-strengthened Corning Gorilla$^\text{\textregistered}$ glass) and in particular to avoid shielding effects. Again, the spatial properties of the raw beam are assumed to be close to the diffraction limit ($M^2<1.3$, fundamental Gaussian mode). An exemplary absorption distribution is shown in \hyperref[fig:scheiss1]{Fig.\,\ref{fig:scheiss1}} [inset shows the computed spot profile, intensity representation, cf. \hyperref[fig:3dfoc]{Fig.\,\ref{fig:3dfoc}\,(c)}] which was recorded using pump-probe microscopy (see Jenne \textit{et al.}\cite{Jenne2019} for experimental details about the diagnostic tool). A single laser burst with $4$ pulses (pulse energy of $\unit[30]{\upmu J}$) emitted from a TruMicro 2000 with a pulse duration of $\unit[5]{ps}$ was used. The measured optical depth $\tau\left(x, z\right)$ (cf. \hyperref[sec:bessel2]{Sec.\,\ref{sec:bessel2}}) at a probe delay of $\unit[5]{ns}$ indicates the optical losses due to, e.g., absorption or scattering on a transient temporal scale. For each of the 13 foci a single absorption zone is observed starting at the geometrical focus and expanding in direction of the incoming laser pulse. The behavior well known from focusing single Gaussian beams into transparent materials\cite{Grossmann2016, Bergner2018} is multiplied here due to the beam splitting concept. Intriguingly, the spatial distribution of induced modification corresponds to the simulated focus positions and even at this small lateral and longitudinal spot separation of $\unit[3]{\upmu m}$ no shielding or inhomogeneities in between the individual spots is obtained. This denotes a precondition for advanced material processing in particular for scaling throughput in the field of cutting, material functionalization or welding. 

\section{Ultrafast drilling and structuring by digital holography}\label{sec:SLM}
\begin{figure}[t]
	\begin{center}
		\includegraphics[width=1.0\columnwidth]{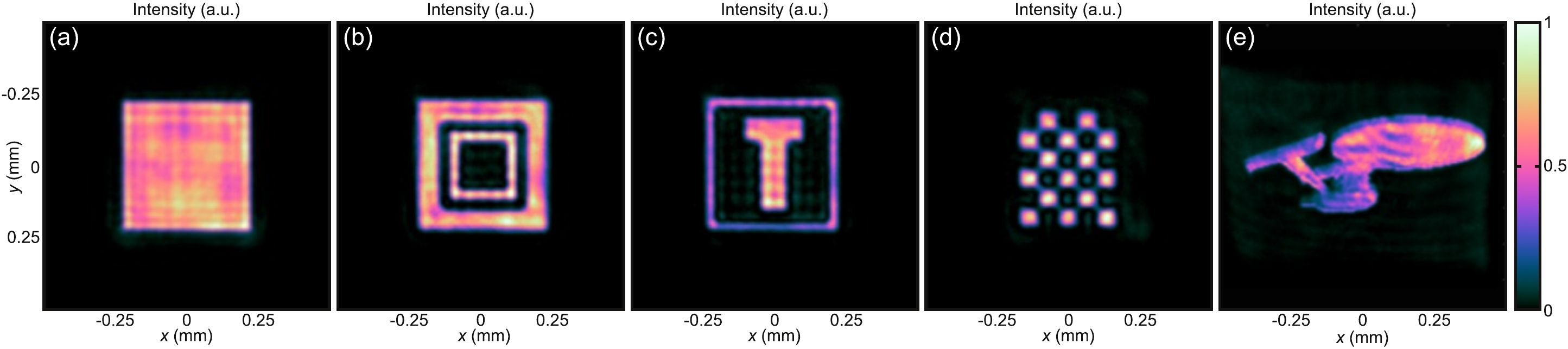}
	\end{center}
\vspace{-0.2cm}
	\caption{Examples of measured beam profiles exhibiting sharp edges and homogeneous intensity distributions for ultrafast drilling and structuring.}
	\label{fig:FMM1}
	\vspace{0.3cm}
\end{figure}

Ultrafast micron-scaled ablation, drilling or marking of solids have become well established industrial processes. Besides the availability of the required light source, one reason for this development can be found in the broad accessibility to multi-axis system technology in combination with advanced scanning systems, allowing arbitrary processing geometries. However, to maximize throughput beam shaping offers unmatched advantages over Gaussian processing. Especially ultrafast microprocesses benefit from large spatial intensity gradients combined with an extended single pulse processing volume. Efficient and flexible diffractive generation of such intensity distributions can be achieved by holographic methods \cite{Dickey2000, Pal2018, Hendriks2012, Naidoo2018}. \hyperref[fig:FMM1]{Figure \ref{fig:FMM1}} shows intensity measurements of exemplary shaped beams. Sharp edges close to the diffraction limit in combination with homogeneous intensity distributions can be applied for structuring or drilling with best quality and maximized throughput. The required micrometer-scaled dimensions are finally achieved using adapted telescopic setups.  

\begin{figure}[t]
	\begin{center}
		\includegraphics[width=1.0\columnwidth]{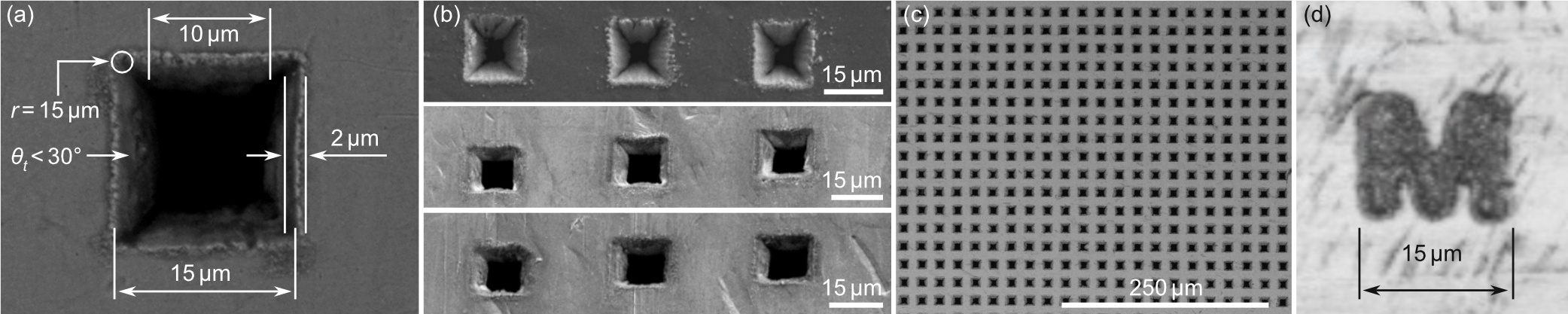}
	\end{center}
\vspace{-0.2cm}
	\caption{Drilling result using the flat-top beam profile depicted in \hyperref[fig:FMM1]{Fig.\,\ref{fig:FMM1}\,(a)}. Direct drilling of quadratic-shaped holes in thin metal sheets with dimensions of $\unit[10]{\upmu m}$ (a), controllable taper geometry and achieved hole density of $\unit[>1000]{ppi}$ (c). Structuring result using a micrometer-scaled ``M''-shaped beam profile (d).}
	\label{fig:guenn1}
\end{figure}

The in \hyperref[fig:FMM1]{Fig.\,\ref{fig:FMM1}\,(a)} depicted flat top intensity profile at $\lambda = \unit[515]{nm}$, generated by a spatial light modulator is demagnified using a $50\times$ $4f$-setup and applied in the following for the direct drilling of micro-shaped-holes. Such optimized drilling processes meet the demand for applications which require flexibility, best quality and high throughput. Such applications may be the drilling of high spatial frequency metal masks for spatially well confined evaporation of substrates for OLED display fabrication. These so-called Fine Metal Masks (FMMs) are subject to special requirements with regard to resolution, geometry and taper angle. For example, rectangular drilling holes with dimensions of about $\unit[10]{\upmu m}$ and resolutions $\unit[>1000]{ppi}$ are required. These level of freedom are easily achieved by means of holographic beam shaping concepts. In \hyperref[fig:guenn1]{Fig.\,\ref{fig:guenn1}} we present as an example processing results of holes with a specifically designed taper angle of $\theta_t = 30^\circ$ and an exit hole-diameter of $\unit[10]{\upmu m}$ achieved by percussion drilling using a TruMicro 2000 Series laser source.

\section{Conclusion}
Structured light concepts applied to ultrafast laser sources were presented and their potential for novel materials processing strategies were discussed. Digital-holographically generated Bessel-like beams with tailored trajectories in all spatial dimensions are used to precisely cut transparent materials with tailored edges. The implementation into industrial cleaving optics was demonstrated allowing single-pass full-thickness modification of glasses with thicknesses of up to $\unit[10]{mm}$. Additionally, we introduced concepts for arbitrarily splitting a large number of single spots into a three-dimensional working volume for exploiting the complete power performance of the laser source -- a precondition for industrial material processing in particular for scaling throughput in the field of cutting, material functionalization or welding. Finally, we made use of digital holography for shaping quadratic micrometer-scaled flat-top-like beams with high-level of homogeneity and sharp edges for ultrafast drilling or structuring applications. This concept was used to generate high spatial frequency metal masks with holes densities of $>\unit[1000]{ppi}$ and controllable taper geometry. 

The potential of structured light concepts was discussed for selected application examples. However, in principle, these concepts could not only be used for a plethora of ultrafast materials processing strategies but also for applications based on continuous-wave lasers, either working in spatial single- or multi-mode regime.



\bibliographystyle{spiebib}   
\bibliography{Lit}

\begin{thebibliography}{10}

\bibitem{Feng1997}
Q.~Feng, J.~V. Moloney, A.~C. Newell, E.~M. Wright, K.~Cook, P.~K. Kennedy,
  D.~Hammer, B.~Rockwell, and C.~Thompson, ``Theory and simulation on the
  threshold of water breakdown induced by focused ultrashort laser pulses,''
  {\em IEEE journal of quantum electronics}~{\bf 33}(2), pp.~127--137, 1997.

\bibitem{Nolte1997}
S.~Nolte, C.~Momma, H.~Jacobs, A.~T{\"u}nnermann, B.~N. Chichkov,
  B.~Wellegehausen, and H.~Welling, ``Ablation of metals by ultrashort laser
  pulses,'' {\em JOSA B}~{\bf 14}(10), pp.~2716--2722, 1997.

\bibitem{Couairon2007}
A.~Couairon and A.~Mysyrowicz, ``Femtosecond filamentation in transparent
  media,'' {\em Physics reports}~{\bf 441}(2-4), pp.~47--189, 2007.

\bibitem{Gattass2008}
R.~R. Gattass and E.~Mazur, ``Femtosecond laser micromachining in transparent
  materials,'' {\em Nature photonics}~{\bf 2}(4), p.~219, 2008.

\bibitem{Shimotsuma2003}
Y.~Shimotsuma, P.~G. Kazansky, J.~Qiu, and K.~Hirao, ``Self-organized
  nanogratings in glass irradiated by ultrashort light pulses,'' {\em Physical
  review letters}~{\bf 91}(24), p.~247405, 2003.

\bibitem{Kumkar2014}
M.~Kumkar, L.~Bauer, S.~Russ, M.~Wendel, J.~Kleiner, D.~Grossmann, K.~Bergner,
  and S.~Nolte, ``Comparison of different processes for separation of glass and
  crystals using ultrashort pulsed lasers,'' in {\em Frontiers in Ultrafast
  Optics: Biomedical, Scientific, and Industrial Applications XIV},   {\bf
  8972}, p.~897214, International Society for Optics and Photonics, 2014.

\bibitem{Mathis2012}
A.~Mathis, F.~Courvoisier, L.~Froehly, L.~Furfaro, M.~Jacquot, P.-A. Lacourt,
  and J.~M. Dudley, ``Micromachining along a curve: Femtosecond laser
  micromachining of curved profiles in diamond and silicon using accelerating
  beams,'' {\em Applied Physics Letters}~{\bf 101}(7), p.~071110, 2012.

\bibitem{Bergner2018b}
K.~Bergner, M.~M{\"u}ller, R.~Klas, J.~Limpert, S.~Nolte, and A.~T{\"u}nnerman,
  ``Scaling ultrashort laser pulse induced glass modifications for cleaving
  applications,'' {\em Applied optics}~{\bf 57}(21), pp.~5941--5947, 2018.

\bibitem{Jenne2018}
M.~Jenne, D.~Flamm, T.~Ouaj, J.~Hellstern, J.~Kleiner, D.~Grossmann,
  M.~Koschig, M.~Kaiser, M.~Kumkar, and S.~Nolte, ``High-quality tailored-edge
  cleaving using aberration-corrected bessel-like beams,'' {\em Optics
  letters}~{\bf 43}(13), pp.~3164--3167, 2018.

\bibitem{Kumkar2017}
M.~Kumkar, M.~Kaiser, J.~Kleiner, D.~Flamm, D.~Grossmann, K.~Bergner,
  F.~Zimmermann, and S.~Nolte, ``Throughput scaling by spatial beam shaping and
  dynamic focusing,'' in {\em Laser Applications in Microelectronic and
  Optoelectronic Manufacturing (LAMOM) XXII},   {\bf 10091}, p.~100910G,
  International Society for Optics and Photonics, 2017.

\bibitem{Bergner2018a}
K.~Bergner, D.~Flamm, M.~Jenne, M.~Kumkar, A.~T{\"u}nnermann, and S.~Nolte,
  ``Time-resolved tomography of ultrafast laser-matter interaction,'' {\em
  Optics express}~{\bf 26}(3), pp.~2873--2883, 2018.

\bibitem{AMPHOSAMPHOS}
{AMPHOS: AMPHOS 400}. \url{https://www.amphos.de/products/amphos-400/}.
\newblock Accessed: 2019-01-07.

\bibitem{Rubinsztein-Dunlop2016}
H.~Rubinsztein-Dunlop, A.~Forbes, M.~V. Berry, M.~R. Dennis, D.~L. Andrews,
  M.~Mansuripur, C.~Denz, C.~Alpmann, P.~Banzer, T.~Bauer, {\em et~al.},
  ``Roadmap on structured light,'' {\em Journal of Optics}~{\bf 19}(1),
  p.~013001, 2016.

\bibitem{Zhu2005}
G.~Zhu, J.~Van~Howe, M.~Durst, W.~Zipfel, and C.~Xu, ``Simultaneous spatial and
  temporal focusing of femtosecond pulses,'' {\em Optics express}~{\bf 13}(6),
  pp.~2153--2159, 2005.

\bibitem{Kammel2014}
R.~Kammel, R.~Ackermann, J.~Thomas, J.~G{\"o}tte, S.~Skupin, A.~T{\"u}nnermann,
  and S.~Nolte, ``Enhancing precision in fs-laser material processing by
  simultaneous spatial and temporal focusing,'' {\em Light: Science \&
  Applications}~{\bf 3}(5), p.~e169, 2014.

\bibitem{Arlt2000}
J.~Arlt, T.~Hitomi, and K.~Dholakia, ``Atom guiding along laguerre-gaussian and
  bessel light beams,'' {\em Applied Physics B}~{\bf 71}(4), pp.~549--554,
  2000.

\bibitem{McGloin2003}
D.~McGloin, V.~Garc{\'e}s-Ch{\'a}vez, and K.~Dholakia, ``Interfering bessel
  beams for optical micromanipulation,'' {\em Optics letters}~{\bf 28}(8),
  pp.~657--659, 2003.

\bibitem{Bhuyan2010}
M.~Bhuyan, F.~Courvoisier, P.~Lacourt, M.~Jacquot, R.~Salut, L.~Furfaro, and
  J.~Dudley, ``High aspect ratio nanochannel machining using single shot
  femtosecond bessel beams,'' {\em Applied Physics Letters}~{\bf 97}(8),
  p.~081102, 2010.

\bibitem{Zhang2018}
G.~Zhang, R.~Stoian, W.~Zhao, and G.~Cheng, ``Femtosecond laser bessel beam
  welding of transparent to non-transparent materials with large focal-position
  tolerant zone,'' {\em Optics express}~{\bf 26}(2), pp.~917--926, 2018.

\bibitem{Durnin1987}
J.~Durnin, J.~Miceli~Jr, and J.~Eberly, ``Diffraction-free beams,'' {\em
  Physical review letters}~{\bf 58}(15), p.~1499, 1987.

\bibitem{McGloin2005}
D.~McGloin and K.~Dholakia, ``Bessel beams: diffraction in a new light,'' {\em
  Contemporary Physics}~{\bf 46}(1), pp.~15--28, 2005.

\bibitem{Flamm2015}
D.~Flamm, D.~Grossmann, M.~Kaiser, J.~Kleiner, M.~Kumkar, K.~Bergner, and
  S.~Nolte, ``Tuning the energy deposition of ultrashort pulses inside
  transparent materials for laser cutting applications,'' {\em Proc. LiM}~{\bf
  253}, 2015.

\bibitem{Dudutis2018}
J.~Dudutis, R.~Stonys, G.~Ra{\v{c}}iukaitis, and P.~Ge{\v{c}}ys,
  ``Aberration-controlled bessel beam processing of glass,'' {\em Optics
  express}~{\bf 26}(3), pp.~3627--3637, 2018.

\bibitem{Leach2006}
J.~Leach, G.~M. Gibson, M.~J. Padgett, E.~Esposito, G.~McConnell, A.~J. Wright,
  and J.~M. Girkin, ``Generation of achromatic bessel beams using a compensated
  spatial light modulator,'' {\em Optics express}~{\bf 14}(12), pp.~5581--5587,
  2006.

\bibitem{AsphericonTechnologies}
{asphericon: Asphericon Technologies}.
  \url{https://www.asphericon.com/en/technologies/}.
\newblock Accessed: 2018-11-20.

\bibitem{PRP}
{Powerphotonic: Rapid Prototyping}.
  \url{http://www.powerphotonic.com/rapid-prototyping}.
\newblock Accessed: 2018-11-26.

\bibitem{Boucher2018}
P.~Boucher, J.~Del~Hoyo, C.~Billet, O.~Pinel, G.~Labroille, and F.~Courvoisier,
  ``Generation of high conical angle bessel--gauss beams with reflective
  axicons,'' {\em Applied optics}~{\bf 57}(23), pp.~6725--6728, 2018.

\bibitem{Brzobohaty2008}
O.~Brzobohat{\`y}, T.~{\v{C}}i{\v{z}}m{\'a}r, and P.~Zem{\'a}nek, ``High
  quality quasi-bessel beam generated by round-tip axicon,'' {\em Optics
  express}~{\bf 16}(17), pp.~12688--12700, 2008.

\bibitem{Kumkar2017a}
M.~Kumkar, J.~Kleiner, D.~Gro{\ss}mann, D.~Flamm, and M.~Kaiser, ``Optical
  system for beam shaping,'' Sept.~14 2017.
\newblock US Patent App. 15/598,816.

\bibitem{Flamm2017}
D.~Flamm, K.~Bergner, D.~Grossmann, J.~Hellstern, J.~Kleiner, M.~Jenne,
  S.~Nolte, and M.~Kumkar, ``Higher-order bessel-like beams for optimized
  ultrafast processing of transparent materials,'' in {\em The European
  Conference on Lasers and Electro-Optics},  p.~CM\_3\_3, Optical Society of
  America, 2017.

\bibitem{Meyer2017}
R.~Meyer, M.~Jacquot, R.~Giust, J.~Safioui, L.~Rapp, L.~Furfaro, P.-A. Lacourt,
  J.~Dudley, and F.~Courvoisier, ``Single-shot ultrafast laser processing of
  high-aspect-ratio nanochannels using elliptical bessel beams,'' {\em Optics
  letters}~{\bf 42}(21), pp.~4307--4310, 2017.

\bibitem{Du2014}
T.~Du, T.~Wang, and F.~Wu, ``Generation of three-dimensional optical bottle
  beams via focused non-diffracting bessel beam using an axicon,'' {\em Optics
  Communications}~{\bf 317}, pp.~24--28, 2014.

\bibitem{Chremmos2012}
I.~D. Chremmos, Z.~Chen, D.~N. Christodoulides, and N.~K. Efremidis,
  ``Bessel-like optical beams with arbitrary trajectories,'' {\em Optics
  letters}~{\bf 37}(23), pp.~5003--5005, 2012.

\bibitem{Green2011}
D.~A. Green, ``A colour scheme for the display of astronomical intensity
  images,'' {\em arXiv preprint arXiv:1108.5083} , 2011.

\bibitem{Kumkar2017b}
M.~Kumkar, J.~Kleiner, D.~Gro{\ss}mann, D.~Flamm, and M.~Kaiser, ``System for
  asymmetric optical beam shaping,'' Sept.~7 2017.
\newblock US Patent App. 15/599,720.

\bibitem{Jenne2018a}
M.~Jenne, D.~Flamm, D.~Grossmann, J.~Hellstern, T.~Ouaj, M.~Kumkar, and
  S.~Nolte, ``Glass cutting optimization with pump-probe microscopy and bessel
  beam profiles,'' in {\em Frontiers in Ultrafast Optics: Biomedical,
  Scientific, and Industrial Applications XVIII},   {\bf 10522}, p.~1052216,
  International Society for Optics and Photonics, 2018.

\bibitem{Dudley2013}
A.~Dudley, M.~Lavery, M.~Padgett, and A.~Forbes, ``Unraveling bessel beams,''
  {\em Optics and Photonics News}~{\bf 24}(6), pp.~22--29, 2013.

\bibitem{Trichili2014}
A.~Trichili, T.~Mhlanga, Y.~Ismail, F.~S. Roux, M.~McLaren, M.~Zghal, and
  A.~Forbes, ``Detection of bessel beams with digital axicons,'' {\em Optics
  express}~{\bf 22}(14), pp.~17553--17560, 2014.

\bibitem{Xie2015}
C.~Xie, V.~Jukna, C.~Mili{\'a}n, R.~Giust, I.~Ouadghiri-Idrissi, T.~Itina,
  J.~M. Dudley, A.~Couairon, and F.~Courvoisier, ``Tubular filamentation for
  laser material processing,'' {\em Scientific reports}~{\bf 5}, p.~8914, 2015.

\bibitem{Jenne2019}
M.~Jenne, D.~Flamm, M.~Faber, D.~Grossmann, J.~Kleiner, F.~Zimmermann,
  M.~Kumkar, and S.~Nolte, ``Pump-probe microscopy of tailored ultrashort laser
  pulses for glass separation processes,'' in {\em Laser-based Micro- and
  Nanoprocessing XIII (to be published)},  International Society for Optics and
  Photonics, 2019.

\bibitem{Grossmann2016}
D.~Grossmann, M.~Reininghaus, C.~Kalupka, M.~Kumkar, and R.~Poprawe,
  ``Transverse pump-probe microscopy of moving breakdown, filamentation and
  self-organized absorption in alkali aluminosilicate glass using ultrashort
  pulse laser,'' {\em Optics express}~{\bf 24}(20), pp.~23221--23231, 2016.

\bibitem{Itoh2006}
K.~Itoh, W.~Watanabe, S.~Nolte, and C.~B. Schaffer, ``Ultrafast processes for
  bulk modification of transparent materials,'' {\em MRS bulletin}~{\bf 31}(8),
  pp.~620--625, 2006.

\bibitem{Rethfeld2004}
B.~Rethfeld, K.~Sokolowski-Tinten, D.~Von Der~Linde, and S.~Anisimov,
  ``Timescales in the response of materials to femtosecond laser excitation,''
  {\em Applied Physics A}~{\bf 79}(4-6), pp.~767--769, 2004.

\bibitem{Wyrowski2015}
F.~Wyrowski and C.~Hellmann, ``Geometrical optics reloaded: Physical optics
  modeling with smart rays,'' {\em Optik \& Photonik}~{\bf 10}(5), pp.~43--47,
  2015.

\bibitem{Noll1976}
R.~J. Noll, ``Zernike polynomials and atmospheric turbulence,'' {\em J. Opt.
  Soc. Am.}~{\bf 66}, pp.~207--211, Mar 1976.

\bibitem{Schulze2013}
C.~Schulze, A.~Dudley, D.~Flamm, M.~Duparr{\'e}, and A.~Forbes,
  ``Reconstruction of laser beam wavefronts based on mode analysis,'' {\em
  Applied optics}~{\bf 52}(21), pp.~5312--5317, 2013.

\bibitem{TRUMPFTOPC}
{TRUMPF: TOP Cleave cutting optics}.
  \url{https://www.trumpf.com/en_US/products/lasers/processing-optics/top-cleave-cutting-optics/}.
\newblock Accessed: 2018-11-28.

\bibitem{Jansen2018}
F.~Jansen, A.~Budnicki, and D.~Sutter, ``Pulsed lasers for industrial
  applications: Fiber, slab and thin-disk: Ultrafast laser technology for every
  application,'' {\em Laser Technik Journal}~{\bf 15}(2), pp.~46--49, 2018.

\bibitem{Kumkar2017c}
M.~Kumkar, D.~Gro{\ss}mann, and D.~Flamm, ``Diffractive optical beam shaping
  element,'' Sept.~28 2017.
\newblock US Patent App. 15/599,623.

\bibitem{Eidam2010}
T.~Eidam, S.~Hanf, E.~Seise, T.~V. Andersen, T.~Gabler, C.~Wirth, T.~Schreiber,
  J.~Limpert, and A.~T{\"u}nnermann, ``Femtosecond fiber {CPA} system emitting
  {830 W} average output power,'' {\em Optics letters}~{\bf 35}(2), pp.~94--96,
  2010.

\bibitem{Fattahi2014}
H.~Fattahi, H.~G. Barros, M.~Gorjan, T.~Nubbemeyer, B.~Alsaif, C.~Y. Teisset,
  M.~Schultze, S.~Prinz, M.~Haefner, M.~Ueffing, {\em et~al.},
  ``Third-generation femtosecond technology,'' {\em Optica}~{\bf 1}(1),
  pp.~45--63, 2014.

\bibitem{Negel2015}
J.-P. Negel, A.~Loescher, A.~Voss, D.~Bauer, D.~Sutter, A.~Killi, M.~A. Ahmed,
  and T.~Graf, ``Ultrafast thin-disk multipass laser amplifier delivering {1.4
  kW} ({4.7 mJ}, {1030 nm}) average power converted to {820 W} at 515 nm and
  {234 W} at 343 nm,'' {\em Optics express}~{\bf 23}(16), pp.~21064--21077,
  2015.

\bibitem{Streibl1989}
N.~Streibl, ``Beam shaping with optical array generators,'' {\em Journal of
  Modern Optics}~{\bf 36}(12), pp.~1559--1573, 1989.

\bibitem{Ren2014}
H.~Ren, H.~Lin, X.~Li, and M.~Gu, ``Three-dimensional parallel recording with a
  debye diffraction-limited and aberration-free volumetric multifocal array,''
  {\em Optics letters}~{\bf 39}(6), pp.~1621--1624, 2014.

\bibitem{Zhu2014}
L.~Zhu, M.~Sun, M.~Zhu, J.~Chen, X.~Gao, W.~Ma, and D.~Zhang,
  ``Three-dimensional shape-controllable focal spot array created by focusing
  vortex beams modulated by multi-value pure-phase grating,'' {\em Optics
  express}~{\bf 22}(18), pp.~21354--21367, 2014.

\bibitem{Gu2014}
M.~Gu, X.~Li, and Y.~Cao, ``Optical storage arrays: a perspective for future
  big data storage,'' {\em Light: Science \& Applications}~{\bf 3}(5), p.~e177,
  2014.

\bibitem{Jesacher2010}
A.~Jesacher and M.~J. Booth, ``Parallel direct laser writing in three
  dimensions with spatially dependent aberration correction,'' {\em Optics
  express}~{\bf 18}(20), pp.~21090--21099, 2010.

\bibitem{Simmonds2011}
R.~D. Simmonds, P.~S. Salter, A.~Jesacher, and M.~J. Booth, ``Three dimensional
  laser microfabrication in diamond using a dual adaptive optics system,'' {\em
  Optics express}~{\bf 19}(24), pp.~24122--24128, 2011.

\bibitem{Flamm2012}
D.~Flamm, D.~Naidoo, C.~Schulze, A.~Forbes, and M.~Duparr{\'e}, ``Mode analysis
  with a spatial light modulator as a correlation filter,'' {\em Optics
  letters}~{\bf 37}(13), pp.~2478--2480, 2012.

\bibitem{Forbes2016}
A.~Forbes, A.~Dudley, and M.~McLaren, ``Creation and detection of optical modes
  with spatial light modulators,'' {\em Advances in Optics and Photonics}~{\bf
  8}(2), pp.~200--227, 2016.

\bibitem{Davis1999}
J.~A. Davis, D.~M. Cottrell, J.~Campos, M.~J. Yzuel, and I.~Moreno, ``Encoding
  amplitude information onto phase-only filters,'' {\em Applied optics}~{\bf
  38}(23), pp.~5004--5013, 1999.

\bibitem{Arrizon2007}
V.~Arriz{\'o}n, U.~Ruiz, R.~Carrada, and L.~A. Gonz{\'a}lez, ``Pixelated phase
  computer holograms for the accurate encoding of scalar complex fields,'' {\em
  JOSA A}~{\bf 24}(11), pp.~3500--3507, 2007.

\bibitem{Leutenegger2006}
M.~Leutenegger, R.~Rao, R.~A. Leitgeb, and T.~Lasser, ``Fast focus field
  calculations,'' {\em Optics express}~{\bf 14}(23), pp.~11277--11291, 2006.

\bibitem{Wyrowski1988}
F.~Wyrowski and O.~Bryngdahl, ``Iterative fourier-transform algorithm applied
  to computer holography,'' {\em JOSA A}~{\bf 5}(7), pp.~1058--1065, 1988.

\bibitem{Jenne2018b}
M.~Jenne, F.~Zimmermann, D.~Flamm, D.~Gro{\ss}mann, J.~Kleiner, M.~Kumkar, and
  S.~Nolte, ``Multi pulse pump-probe diagnostics for development of advanced
  transparent materials processing,'' {\em JLMN}~{\bf 13}(3), pp.~273--279,
  2018.

\bibitem{Bergner2018}
K.~Bergner, B.~Seyfarth, K.~Lammers, T.~Ullsperger, S.~D{\"o}ring, M.~Heinrich,
  M.~Kumkar, D.~Flamm, A.~T{\"u}nnermann, and S.~Nolte, ``Spatio-temporal
  analysis of glass volume processing using ultrashort laser pulses,'' {\em
  Applied optics}~{\bf 57}(16), pp.~4618--4632, 2018.

\bibitem{Dickey2000}
F.~M. Dickey and S.~C. Holswade, ``Laser beam shaping: Theory and techniques,''
  2000.

\bibitem{Pal2018}
V.~Pal, C.~Tradonsky, R.~Chriki, N.~Kaplan, A.~Brodsky, M.~Attia, N.~Davidson,
  and A.~A. Friesem, ``Generating flat-top beams with extended depth of
  focus,'' {\em Applied optics}~{\bf 57}(16), pp.~4583--4589, 2018.

\bibitem{Hendriks2012}
A.~Hendriks, D.~Naidoo, F.~S. Roux, C.~L{\'o}pez-Mariscal, and A.~Forbes, ``The
  generation of flat-top beams by complex amplitude modulation with a
  phase-only spatial light modulator,'' in {\em Laser Beam Shaping XIII},
  {\bf 8490}, p.~849006, International Society for Optics and Photonics, 2012.

\bibitem{Naidoo2018}
D.~Naidoo, I.~A. Litvin, and A.~Forbes, ``Brightness enhancement in a
  solid-state laser by mode transformation,'' {\em Optica}~{\bf 5}(7),
  pp.~836--843, 2018.

\end{thebibliography}

\end{document}